\documentclass[%
 reprint,
superscriptaddress,
groupedaddress,
aps,superscriptaddress,
pra,
]{revtex4-2}
\usepackage{graphicx,braket,color,bm,amsmath,amssymb}
\begin{document}

\title{Mott Criticality as the Confinement Transition of a Pseudogap-Mott Metal}

\author{Abhirup Mukherjee}
\affiliation{Department of Physical Sciences, Indian Institute of Science Education and Research Kolkata, Nadia - 741246, India
}
\author{S. R. Hassan}
\affiliation{The Institute of Mathematical Sciences, HBNI, C.I.T. Campus, Chennai 600 113, India}
\author{Anamitra Mukherjee}
\affiliation{School of Physical Sciences, National Institute of Science, Education and Research, HBNI, Jatni 752050, India}
\author{N. S. Vidhyadhiraja}
\affiliation{Theoretical Sciences Unit, Jawaharlal Nehru Center for Advanced Scientific Research, Jakkur, Bengaluru 560064, India}
\author{A. Taraphder}
\affiliation{Department of Physics, Indian Institute of Technology Kharagpur, Kharagpur 721302, India}

\author{Siddhartha Lal}
\affiliation{Department of Physical Sciences, Indian Institute of Science Education and Research Kolkata, Nadia - 741246, India
}

\date{\today}

\begin{abstract}
\noindent
The phenomenon of Mott insulation involves the localization of itinerant electrons due to strong local repulsion. Upon doping, a pseudogap (PG) phase emerges - marked by selective gapping of the Fermi surface without conventional symmetry breaking in spin or charge channels. A key challenge is understanding how quasiparticle breakdown in the Fermi liquid gives rise to this enigmatic state, and how it connects to both the Mott insulating and superconducting phases. Here, we develop a renormalization-based construction of strongly correlated lattice models that captures the emergence of the pseudogap phase and its transition to a Mott insulator. Applying a many-body tiling scheme to the fixed-point impurity model uncovers a lattice model with electron interactions and Kondo physics. At half-filling, the interplay between Kondo screening and bath charge fluctuations in the impurity model leads to Fermi liquid breakdown. This reveals a pseudogap phase characterized by a non-Fermi liquid (the Mott metal) residing on nodal arcs, gapped antinodal regions of the Fermi surface, and an anomalous scaling of the electronic scattering rate with frequency. The eventual confinement of holon–doublon excitations of this exotic metal obtains a continuous transition into the Mott insulator. Our results identify the pseudogap as a distinct long-range entangled quantum phase, and offer a new route to Mott criticality beyond the paradigm of local quantum criticality. 

\end{abstract}

\maketitle
\par\noindent\textbf{\large Introduction}\\
Understanding the breakdown of Fermi liquid (FL) behavior and the origin of the pseudogap (PG) in doped Mott systems remains a central challenge in correlated electron physics. The PG phase – characterised by a selective depletion of low-energy spectral weight on the Fermi surface – has been widely reported in experiments~\cite{taillefer2010,FradkinRevModPhys2015,Hashimoto2014,Sato2017,DoironLeyraud2017,Mukhopadhyay2019,Naman2021,Masafumi2025} and studied theoretically within the hole-doped two-dimensional Hubbard model~\cite{KyungKotliar2006,MacridinAzevedo2006,sakai2009evolution,werner2009momentum,sakai2010doped,lin2010,gull2010momentum,gull2012,Mirzaei2013,gull2013,anirbanmott2,HilleAndergassen2020,Jiang2022}. Yet its microscopic origin, connection to Mott and superconducting phases~\cite{Scheurer2018,Krien2022,Kitatani2023,Sorella2023}, and nature of the nodal–antinodal dichotomy~\cite{Schafer2021} remain unresolved. Furthermore, the origin of experimentally observed finite-temperature crossovers in $T=0$ quantum states is poorly understood~\cite{White1998,Ido2018,ProustTaillefer2019,Ponsioen2019,XuZhang2022}.

\noindent
The breakdown of Landau quasiparticles in the PG regime is tied to the emergence of Luttinger surfaces - zero contours of the Green’s function that fragment the Fermi surface (FS) into nodal arcs~\cite{Phillips2013,dzyaloshinskii2003some}. These surfaces proliferate with increasing interaction strength, reconfiguring the Fermi volume~\cite{seki2017topological}, and breaking the emergent $\mathbb{Z}_2$ symmetry associated with separately conserved spin currents in the FL~\cite{Anderson2001,Huang2022}. Unlike conventional instabilities, these zeros signal a fundamental topological obstruction to quasiparticle formation~\cite{Coleman_2001,Phillips2013Unparticles}. Yet the microscopic mechanism behind their formation, the stability of the resulting non-Fermi liquid (NFL) metal, and the transition to the Mott insulating state remain poorly understood.

\begin{figure*}[tbh]
    \centering
    \includegraphics[width=\linewidth]{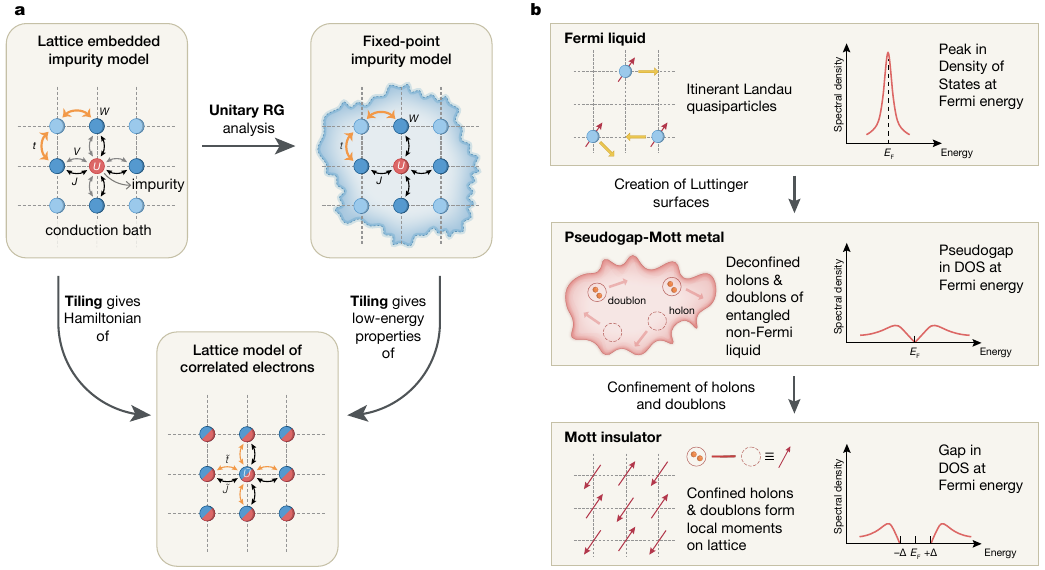}
    \caption{Schematic representation of (a): flowchart of theoretical framework adopted by us (see main text for explanation of symbols), (b): phases obtained for the extended Hubbard model in two dimensions at $T=0$, and passage between them.}
    \label{schematic}
\end{figure*}

\noindent
We address these questions by developing a renormalization group (RG)-based impurity-to-lattice framework that reconstructs the low-energy physics of a strongly correlated system from the fixed-point structure of a quantum impurity model (Fig.~\ref{schematic}, left). Starting from a dynamical Anderson impurity embedded in a correlated bath, we solve the system using a unitary RG (URG) approach~\cite{anirbanurg1,anirbanurg2}. We then construct a translation-invariant lattice Hamiltonian through a many-body tiling procedure that preserves non-local correlations and inherits momentum-resolved self-energies and renormalized couplings from the impurity solution. The resulting model is an \textit{extended}-Hubbard Hamiltonian on a two-dimensional square lattice at half-filling, with both on-site repulsion and nearest-neighbour spin exchange interactions. These correlations, derived directly from the impurity dynamics, enable tracking the evolution from the FL to the Mott insulator (MI) via an intermediate PG phase.

\noindent
A central insight of our work is that the PG phase hosts a NFL state governed by Kondo frustration and doublon–holon deconfinement~\cite{keimer2015quantum} (Fig.~\ref{schematic}, right). A correlation scale ($W$) in the conduction bath suppresses Kondo screening (with coupling $J$) beyond a critical ratio of $W/J$, triggering the emergence of Luttinger surfaces in the lattice model. This breakdown of quasiparticles begins at the antinodes and progressively engulfs the nodal regions, driving a continuous transformation into the MI. The resulting NFL exhibits a pseudogapped density of states, multipartite long-range entanglement, and a universal scattering rate of the form $\sim (a + b\,\omega^2)^{-1}$ that grows as $\omega \to 0$. At the critical endpoint of this phase, the URG flow identifies a singular NFL described by the Hatsugai–Kohmoto (HK) model~\cite{Hatsugai1992,Baskaran1991,Huang2022}, with fully deconfined holon–doublon excitations.

\noindent
The emergence of Luttinger surfaces alters the anomaly structure associated with the generalized symmetry of the FS~\cite{Altshuler_1998,Heath_2020,lanave2025}. The corresponding topological charge is defined via the Luttinger–Ward functional. This charge is modified by Green’s function zeros, signalling a shift in the underlying anomaly. Within our framework, this reorganization grants topological protection to the RG flows terminating in the PG regime, ensuring the stability of its NFL excitations. These are adiabatically connected to the Hatsugai–Kohmoto model~\cite{Hatsugai1992,Baskaran1991}, reinforcing the interpretation of the PG as a distinct quantum phase. We thus identify the NFL state as a new gapless, long-range entangled phase of correlated matter - the \textit{Mott metal} - whose deconfined holon-doublon excitations are eventually confined across the Mott transition~\cite{Mott_1949}.

\noindent
While we demonstrate our approach for an extended Hubbard model, the underlying framework is broadly applicable to correlated quantum systems. The impurity-plus-tiling construction accommodates multi-orbital degrees of freedom, frustrated geometries, and nontrivial band topology. By deriving lattice Hamiltonians directly from renormalized local dynamics - without relying on mean-field approximations - this method preserves nonlocal correlations and enhances momentum-space resolution. It thus provides a versatile and controlled route for engineering strongly correlated models, particularly in regimes where PG formation, NFL scaling, and Mott physics intersect.

\begin{figure*}
    \centering
    \includegraphics[width=\textwidth, height=10cm]{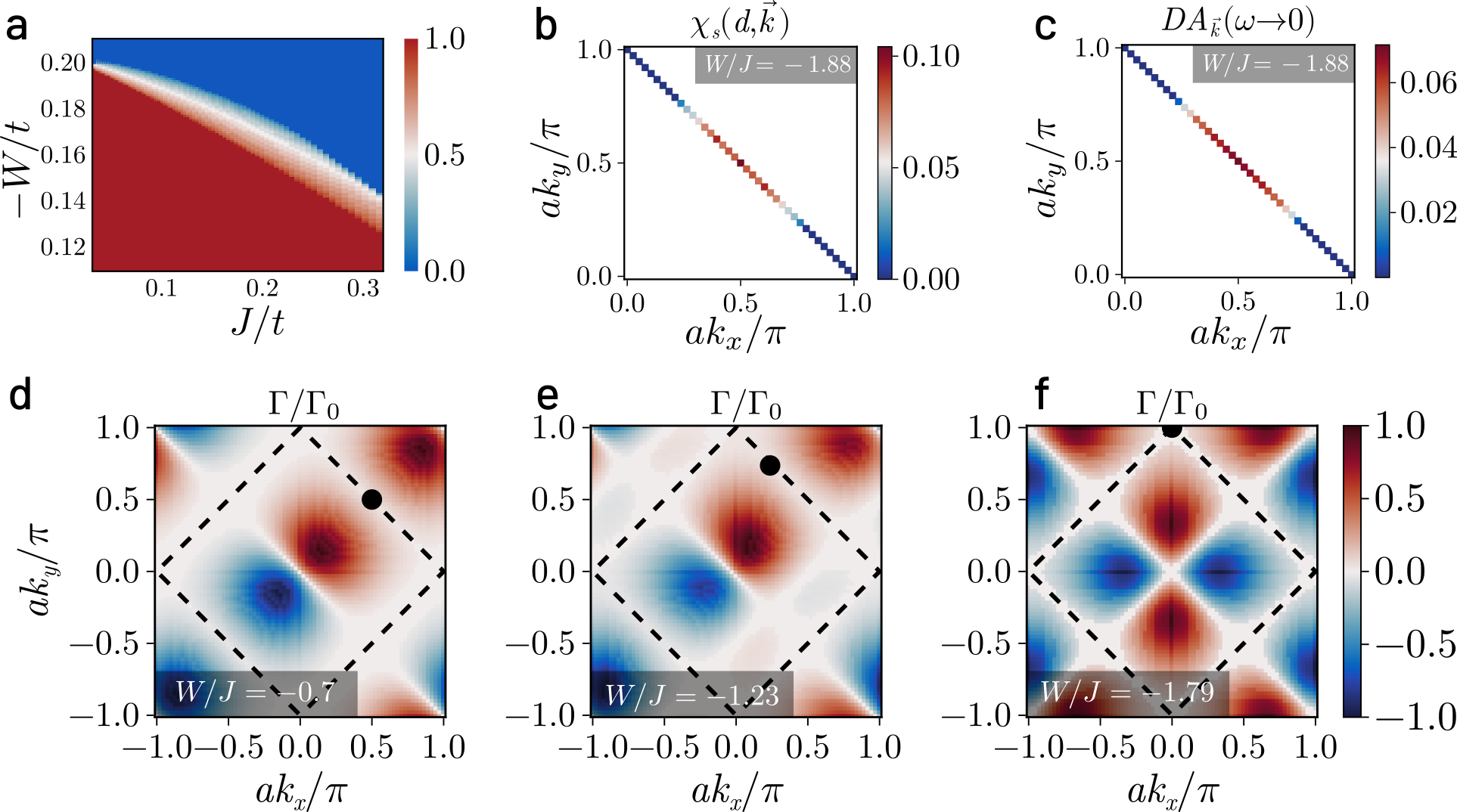}
    \caption{(a): Phase diagram of impurity model at strong coupling in $U$ in terms of competing dimensionless Kondo ($J/t$) and bath correlation ($W/t$) couplings. Colourbar represents the fraction of states on the Fermi surface replaced by zeros of the Greens function (Luttinger surface). A pseudogap phase (shaded region) is observed between the local Fermi liquid (red region) and local moment (blue region) phases (b): $k$-space-resolved spin-spin correlation $\chi_s(d,{\bf k}) = \braket{{\bf S}_d\cdot{\bf S}_{\bf k}}$ in the pseudogap phase of the impurity model. Antinodal regions are observed to decouple from Kondo screening of the impurity. (c): Upon tiling, this leads to a 
    $k$-space-resolved antinodal gap in the electronic density of states of the lattice model, corresponding to Luttinger surfaces of zeros. (d)-(f): Initiation of the decoupling of $J_{{\bf k}, {\bf k'}}$ (positive, negative and zeros shown in red, blue and white respectively), with ${\bf k}$ (black circle) at low-energies for ${\bf k}$ corresponding to the node, antinode and a point mid-way between them on the top right arm of the FS respectively with tuning $W/J$. As dictated by the symmetry of $J_{{\bf k}, {\bf k'}}$, the decoupling for a given ${\bf k}$ proceeds via the appearance of zeros (white patches) of $J_{{\bf k}, {\bf k'}}$ for ${\bf k'}$ initially on the nodal regions of adjacent arms, and progresses gradually towards the antinodes. The decoupling ends with the onset of the pseudogap.}
    \label{phaseDiagram}
\end{figure*}

\vspace{0.25cm}

\begin{figure*}[tbh]
    \centering
    \includegraphics[width=\textwidth]{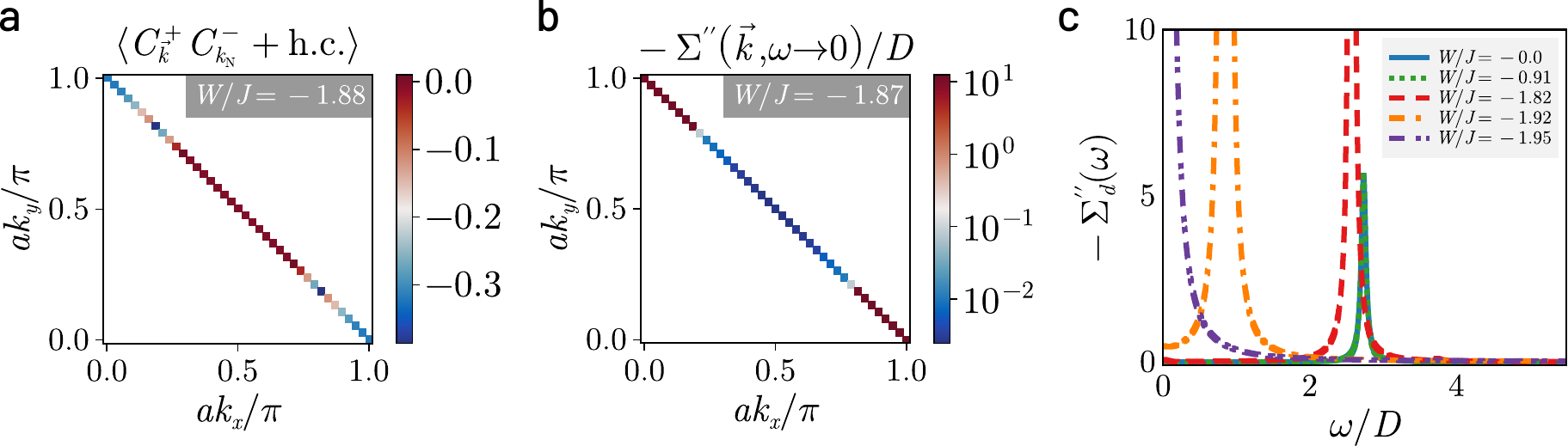}
    \caption{(a): Enhanced charge correlations $\chi_c({\bf k}_1, {\bf k}_2) = \braket{c^\dagger_{{\bf k}_1\uparrow}c^\dagger_{{\bf k}_1\downarrow}c_{{\bf k}_2\downarrow}c_{{\bf k}_2\uparrow} + \text{h.c.}}$ between the nodal and antinodal regions, signalling Kondo breakdown in the pseudogap phase of the impurity model. (b): In turn, the breakdown leads to the gapping of the antinodal regions in the lattice model, seen from the appearance of poles in the imaginary part of the self-energy. (c): The imaginary part of the impurity self-energy $\Sigma^{\prime\prime}(\omega >0)$ possesses a pole at non-zero $\omega$ in the PG that moves towards $\omega=0$ as the Mott transition is approached.}
    \label{chargeCorr}
\end{figure*}

\vspace{0.25cm}
\par\noindent\textbf{\large Tiling Reconstruction of Lattice Model}\\ 
To capture the interplay between Kondo physics and Mott localization, our approach relies on a two-dimensional auxiliary model inspired by insights on the impurity model~\cite{Mukherjee_2023} at the heart of dynamical mean-field theory (DMFT) ~\cite{georges1996} and its extensions ~ \cite{hettler1998,kotliar2001,maier2005}. The model (see Methods for details) consists of a correlated impurity site (with a double occupancy cost $U$) embedded within a correlated conduction bath defined on a square lattice (see Fig.~\ref{schematic}(a)). The impurity site hybridises with the conduction bath sites adjacent to it through one-particle hybridisation $V$ as well as a spin-exchange interaction $J$ arising from fluctuations in local spin densities of the electrons. The conduction bath has minimal correlations in the form of a local correlation term $W$ on the sites that are adjacent to the impurity. Crucially, the Kondo coupling acquires a momentum structure upon Fourier transforming:
\begin{equation}
J_{{\bf k}, {\bf k}^\prime} = \frac{J}{2} \left[ \cos(k_x - k^\prime_x) + \cos(k_y - k^\prime_y) \right],
\end{equation}
which respects the $C_4$ symmetry of the square lattice.

\noindent
To reconstruct a translationally invariant lattice Hamiltonian from the impurity model (with Hamiltonian $\mathcal{H}_\text{aux}$), we define a many-body tiling prescription that embeds the auxiliary impurity system uniformly across the two-dimensional square lattice. Thus,  the tiling process systematically derives the full lattice dynamics from the renormalized fixed-point structure of the impurity. Let ${\bf r}_d$ denote the location of the impurity within a single unit cell. We define the full Hamiltonian by translating this unit cell across all sites of the lattice:
\begin{equation}
\mathcal{H}_\text{tiled} = \sum_{{\bf r}} T^\dagger({\bf r}) \, \mathcal{H}_\text{aux}({\bf r}_d) \, T({\bf r}) - N \mathcal{H}_\text{cbath},
\end{equation}
where $T({\bf r})$ denotes the operator that translates all degrees of freedom by vector ${\bf r}$, and $N$ is the number of lattice sites. Subtracting $N \mathcal{H}_\text{cbath}$ ensures proper normalization and avoids double-counting of bath terms.

\noindent
This procedure yields an effective \textit{extended}-Hubbard model that retains local and nonlocal correlation effects from the original impurity. The resulting Hamiltonian reads:
\begin{equation}
\label{eq:effectiveH}
\begin{aligned}
\hspace*{-0.3cm}\mathcal{H}_\text{tiled} =
&- \frac{\tilde{t}}{\sqrt{\mathcal{Z}}} \sum_{\langle {\bf r}_i, {\bf r}_j \rangle, \sigma} \left( c^\dagger_{{\bf r}_i,\sigma} c_{{\bf r}_j,\sigma} + \text{h.c.} \right) \\
%- \tilde{\mu} \sum_{{\bf r},\sigma} \hat{n}_{{\bf r},\sigma} \\
&+ \frac{\tilde{J}}{\mathcal{Z}} \sum_{\langle {\bf r}_i, {\bf r}_j \rangle} {\bf S}_{{\bf r}_i} \cdot {\bf S}_{{\bf r}_j}
- \frac{\tilde{U}}{2} \sum_{{\bf r}} \left( \hat{n}_{{\bf r}, \uparrow} - \hat{n}_{{\bf r}, \downarrow} \right)^2,
\end{aligned}
\end{equation}
where $\mathcal{Z} = 4$ is the coordination number of the square lattice. The parameters $(\tilde{t}, %\tilde{\mu}, 
\tilde{J}, \tilde{U})$ are renormalized couplings inherited from the impurity solution: $\tilde{t} = t + 2V,\quad \tilde{U} = U + W,
%\quad \tilde{\mu} = 2\mu + \eta,
\quad \tilde{J} = 2J$. A non-zero chemical potential in the bath and/or the impurity can be included to study the effects of doping. Crucially, the eigenstates of $\mathcal{H}_\text{tiled}$ obey a many-body generalization of Bloch’s theorem~\cite{stoyanova} (see Sec.1 of Supplementary Information~\cite{suppmat}), which allows for the exact computation of momentum-resolved observables (presented here for a $77\times 77$ {\bf k}-space Brillouin zone grid). This lattice embedding provides a controlled route to capture low-energy NFL behavior and Mott criticality directly from a quantum impurity model.

\vspace{0.25cm}
\par\noindent\textbf{\large  Pseudogap Formation via Kondo Breakdown}\\
A detailed picture of the PG in the impurity model is obtained from a momentum-resolved breakdown of Kondo screening in the strong-coupling regime $U \gg t$ phase of $H_\text{aux}$. A unitary renormalisation group (URG) scaling analysis~\cite{anirbanurg1} obtains a flow equation of the Kondo coupling \(J^{(j)}_{{\bf k}_1, {\bf k}_2}\) (see Sec.2 of~\cite{suppmat})
\begin{equation}\begin{aligned}\label{KondoRGequation}
	\Delta J^{(j)}_{{\bf k}_1, {\bf k}_2} = -\sum_{{\bf q} \in \text{PS}} \frac{J^{(j)}_{{\bf k}_2,{\bf q}} J^{(j)}_{{\bf q},{\bf k}_1} + 4J^{(j)}_{{\bf q}, {\bf \bar q}} W_{{\bf \bar q}, {\bf k}_2, {\bf k}_1, {\bf q}}}{\omega - \frac{1}{2}|\varepsilon_j| + J^{(j)}_{{\bf q}}/4 + W_{{\bf q}}/2}~,
\end{aligned}\end{equation}
where \(\varepsilon_j\) is the energy of the shell being decoupled at the \(j^\text{th}\) step, the sum is over all occupied momentum states \({\bf q}\) of the energy shell \(\varepsilon_j\), and  \({\bf \bar q} = {\bf q} + {\boldsymbol \pi}\) is the particle-hole transformed state associated with ${\bf q}$. The bath interaction coupling $W_{{\bf \bar q}, {\bf k}_2, {\bf k}_1, {\bf q}}$ is found to be marginal under these transformations. While we present a detailed numerical evaluation of the RG equation for $J_{{\bf k}_1, {\bf k}_2}$ (eq.\eqref{KondoRGequation}) below, it is clear that the frustration of Kondo screening due to charge fluctuations (for attractive bath interactions $W<0$) leads to the Mott transition~\cite{Mukherjee_2023}.

\noindent
Upon tuning the ratio of the bath and Kondo interactions ($W/J$) from zero to negative values (see phase diagram in Fig.~\ref{phaseDiagram}(a)), the following phases emerge in the impurity model from the competition between $J$ and $W$ in eq.~\eqref{KondoRGequation}: (i) for $W/J<(W/J)_{\text{PG}}$, an LFL phase (red region), where the entire FS participates in Kondo screening, (ii) for $\frac{W}{J} \in [(\frac{W}{J})_{\text{PG}}, (\frac{W}{J})_c]$ (shaded region), a local PG phase where disconnected parts of the FS around the node participate in Kondo screening, and (iii) a local moment phase for $\frac{W}{J} > (\frac{W}{J})_c$ (blue region), where the impurity remains unscreened at low-energies. These can be visualised from spin correlations, $\chi_s(d,{\bf k}) = \braket{{\bf S}_d\cdot{\bf S}_{\bf k}}$~, as shown in Fig.~\ref{phaseDiagram}(b) for the PG. The values $(W/J)_{\text{PG}}$ and $(W/J)_c$ are therefore the entry into and exit from the PG phase. Mapping onto the lattice model via tiling (see Sec.3 of \cite{suppmat}), we observe that RG-induced nodal–antinodal dichotomy in $J_{{\bf k}, {\bf k}^\prime}$ in the impurity model is the microscopic origin of the PG in the lattice model: the extinction of Kondo coherence translates into spectral zeros in the lattice Green’s function (i.e., Luttinger surfaces) at antinodal momenta (Fig.~\ref{phaseDiagram}(c)). This establishes that the $T=0$ Mott transition of the 2D \textit{extended}-Hubbard model proceeds from FL to MI through an intervening PG phase.

\begin{figure*}
    \centering
    \includegraphics[width=\textwidth]{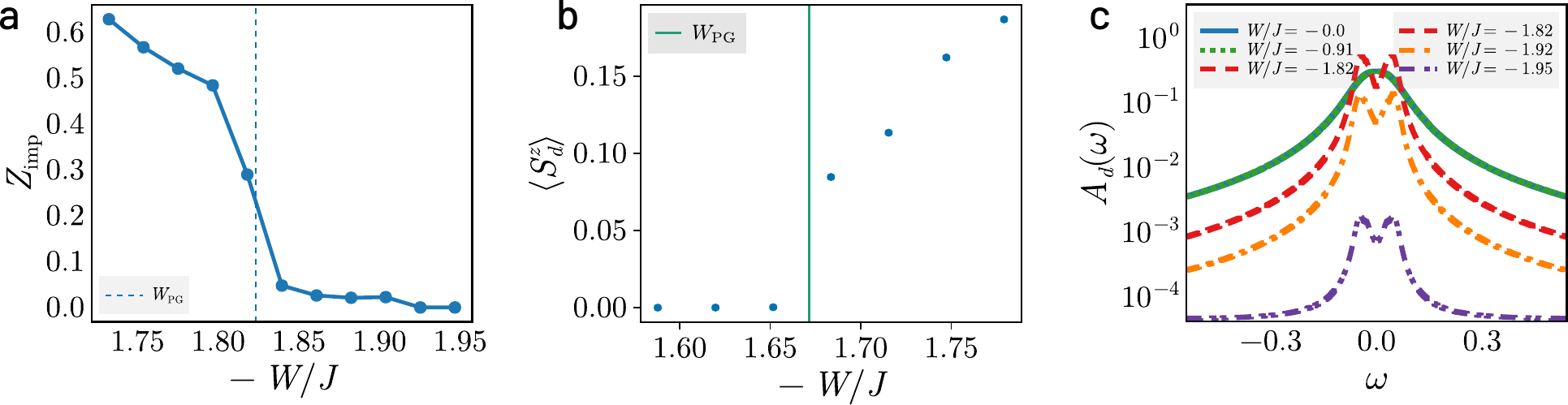}
    \caption{(a): Suppression of quasiparticle residue ($Z_\text{imp}$) as the impurity model is tuned towards the Mott transition. An initial drastic fall in $Z_\text{imp}$ is observed for $W/J \lesssim (W/J)_\text{PG}$ from $0.3$ to around $0.05$, signalling the destruction of the FL with unravelling of Kondo screening. A steady decrease in $Z_{imp}$ is observed in passage through the PG, and is vanishingly small close to the Mott transition due to a divergent self-energy. (b): Growth of unscreened impurity magnetic moment in the pseudogap phase, signalling the breakdown of Kondo screening. (c): The central (Kondo) resonance of the impurity spectral function in the Fermi liquid phase splits into a pseudogap in the PG phase, with its height diminishing rapidly as the Mott transition is approached.
    }
    \label{channelDecoupling}
\end{figure*}
\vspace{0.25cm}
\par\noindent\textbf{\large Unravelling of Kondo Screening}\\
The ${\bf k}$-space anisotropy of Kondo breakdown can be visualized in terms of zeros of $J_{{\bf k}_N, {\bf k}}$, involving spin-flip scattering between the node ${\bf k}_N = (\pi/2, \pi/2)$  and a general wavevector ${\bf k}$.
For any $W/J$, the $\mathcal{C}_{4}$ lattice symmetry dictates that $J_{{\bf k}_N, {\bf k}}$ vanishes if ${\bf k}$ belongs to any of the antinodes or adjacent nodes. Tuning $W/J$ towards $(W/J)_{\text{PG}}$ leads to an unravelling of the Kondo screening: $J_{{\bf k}_N, {\bf k}}$ for ${\bf k}$ close to the adjacent nodes turns RG-irrelevant first, and a patch of zeros subsequently appears in $J_{{\bf k}_N, {\bf k}}$ around this point (Fig.~\ref{spinCorr} (d)). Tuning $W/J$ further extends the patch of zeros towards the antinodes (Fig.~\ref{spinCorr} (e) and (f)). Kondo screening thus unravels by a systematic decoupling of all $J_{{\bf k}_1, {\bf k}_2}$ that connect adjacent quadrants of the Brillouin zone.   
Precisely at $W/J=(W/J)_{\text{PG}}$, the antinode joins this connected region of zeros in $J_{{\bf k}_1, {\bf k}_2}$, marking the decoupling of the antinodes from all other points in the neighbourhood of the FS. This is an interaction-driven Lifshitz transition of the FS, and marks the entry into a PG phase possessing Fermi arcs~\cite{WuFerrero2018}. Importantly, it coincides with an emergent two-channel Kondo (2CK) impurity model, where each channel corresponds to a pair of Fermi arcs on opposite faces of the conduction bath FS. The 2CK nature of the PG is guaranteed by the symmetry of $J_{{\bf k},{\bf k'}}$: $J_{{\bf k},{\bf k'}}= -J_{{\bf k}+{\bf Q},{\bf k'}}=-J_{{\bf k},{\bf k'}+{\bf Q}}$, where ${\bf Q} = (\pi,\pi)$. The PG expands by shrinking these disconnected Fermi arcs towards the respective nodes, leading to nodal metals whose disappearance heralds the Mott transition. 

\vspace{0.25cm}
\par\noindent\textbf{\large Momentum-resolved Dynamical Spectral Weight Transfer}\\
Passage through the PG phase is accompanied by a highly structured transfer of spectral weight across the FS. Strong charge fluctuations develop between the nodal and antinodal regions of the FS in the PG regime of the impurity model (Fig.~\ref{chargeCorr} (a)), as captured by the correlator:
\begin{equation}
\chi_c({\bf k}_1, {\bf k}2) = \left\langle
c^{\dagger}_{{\bf k}1 \uparrow} c^{\dagger}_{{\bf k}1 \downarrow} c_{{\bf k}2 \downarrow} c_{{\bf k}_2 \uparrow} + \text{h.c.}\right\rangle~.
\end{equation}
These fluctuations dynamically redistribute low-energy spectral weight from the antinodes to higher energies, leading to selective gap formation. Accordingly, the Luttinger surfaces of the PG~\cite{dzyaloshinskii2003some} coincides with the appearance of poles of the lattice model self-energy $\Sigma ({\bf k},\omega=0)$ near the antinodes; these poles approach the nodes on tuning towards the Mott transition (Fig.~\ref{chargeCorr} (b)). This mirrors the coalescing of finite-frequency poles of the self-energy poles towards zero frequency in the underlying impurity model~(Fig.~\ref{chargeCorr} (c)). 

\begin{figure*}
    \centering
    \includegraphics[width=\textwidth]{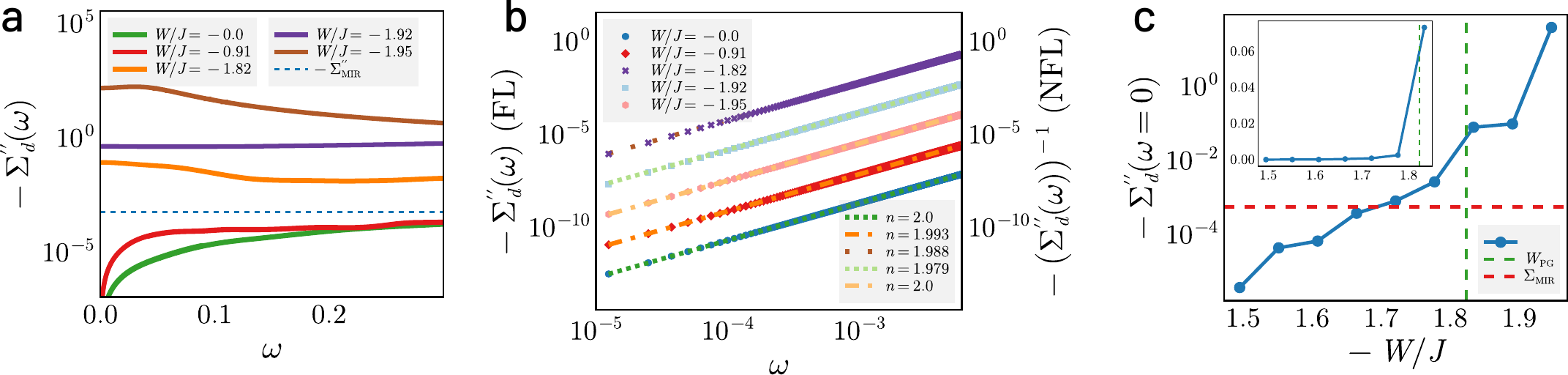}
    \caption{(a): Imaginary part of impurity self-energy $\Sigma^{\prime\prime}(\omega)$ as a function of frequency $\omega$ for Fermi liquid (FL, $|W/J| < 1.79$) and pseudogapped phases (NFL, $|W/J| \geq 1.79$). $\Sigma''(\omega)$ falls to zero as $\omega\to 0$ for the FL, while it attains a peak in the pseudogap. All $\Sigma''(\omega)$ for the NFL are observed to lie above the Mott-Ioffe-Regel (MIR) bound (dashed blue line), while those for the FL lie below. (b): Scaling of $\Sigma^{\prime\prime}(\omega)$ with frequency for Fermi liquid and pseudogapped phases. The FL self-energy fits to $\Sigma^{\prime\prime} \sim \omega^{\alpha}$ with $\alpha\approx 2$, vanishing as $\omega\to 0$, while the NFL self-energy grows as $\Sigma^{\prime\prime} \approx a + b\omega^{\beta}$ for small $\omega$, with $\beta\approx 2$. Remarkably, the NFL exponent remains mostly unchanged through the entirety of the PG phase. (c): Variation of the zero-frequency imaginary self-energy $-\Sigma''(\omega=0)$ with $\omega$ in the FL and pseudogap phases (entry into the PG is marked by the vertical dashed line). The inset shows the same but in linear scale, in order to display the dramatic rise (by almost 30 times) on entering the PG.}
    \label{selfEnergy}
\end{figure*}

\vspace{0.25cm}
\par\noindent\textbf{\large Non-Fermi liquid excitations within the Pseudogap}\\
In the PG regime, the nature of gapless Fermi arcs changes dramatically. We have already argued that the low-energy dynamics of these gapless Fermi arcs are governed by an underlying two-channel Kondo (2CK) impurity model~\cite{Tsvelick_weigmann_mchannel_1985, emery_kivelson}. This is consistent with the rapid fall of the impurity quasiparticle residue $Z_\text{imp}$ (Fig.~\ref{channelDecoupling} (a)) from finite values in the FL phase to vanishingly small values just before the onset of the PG. Fig.~\ref{channelDecoupling} (b) shows the simultaneous emergence of increasingly uncompensated local magnetic moments upon traversing the PG phase. The accompanying impurity spectral function of the gapless arcs show a pseudogapped behaviour for $\omega\simeq 0$, with a rapid fall in the spectral weight at $\omega=0$ upon traversing the PG~(Fig.~\ref{channelDecoupling} (c)). The collapse of the Kondo resonance into a pseudogapped spectral function is accompanied by the redistribution of spectral weight~\cite{dzyaloshinskii2003some} in the impurity spectral function from $\omega\sim 0$ to the Hubbard sidebands at finite frequencies $\omega\simeq \pm 3$ (in units of the bandwidth) (Fig.4 of \cite{suppmat}). Concomitant with this is the emergence of a zero-frequency peak in the imaginary part of the self-energy of the NFL in the PG phase, $-\Sigma''(\omega)\sim (a + \omega^{\beta})^{-1}$~, with $a$ being a constant~(Fig.\ref{selfEnergy} (a)). Remarkably, we find that the exponent $\beta=2$ characterises the NFL for the entire PG phase (see Fig.~\ref{selfEnergy}(b)), including the critical end-point. This is in stark contrast with the $\Sigma''(\omega)\sim \omega^{2}$ for the FL (Fig.~\ref{selfEnergy}(b)). 

\noindent
All $\Sigma''(\omega)$ for the NFL are observed to lie above the Mott-Ioffe-Regel (MIR) bound~\cite{GunnarssonRMP2003,Hussey2004} (dashed blue line in Fig.\ref{selfEnergy} (a)), while those for the FL lie below. The MIR bound is the maximum expected scattering rate in metals when the mean free path approaches the lattice spacing: 
$-2\Sigma^{\prime\prime}_\text{MIR} = 1/\tau_\text{MIR}~, ~\tau_\text{MIR} = l_\text{min}/v_{F}$~,where $l_\text{min}$ is the minimum mean free path in metals (equal to one lattice spacing) and $\tau_\text{MIR}$ is the associated lifetime of quasiparticles close to the FS (with Fermi velocity $v_F$). The optical conductivity $\sigma (\omega)$ of the NFL thus undergoes a suppression as $\omega \to 0$, together with a shift of the Drude peak to finite $\omega$~\cite{Hussey2004,pustogow2021}. The height of the zero-frequency peak rises by almost 4 orders of magnitude from the start of the PG till its end at the Mott transition point (Fig.\ref{selfEnergy}(c)); the dramatic growth of the peak height very near the Mott critical point coincides with the coalescing of the finite-frequency poles of the self-energy into a single pole at zero-frequency, signalling the singular nodal NFL present at the Mott quantum critical point.

\vspace{0.25cm}
\par
\begin{figure*}
    \centering
    \includegraphics[width=\textwidth]{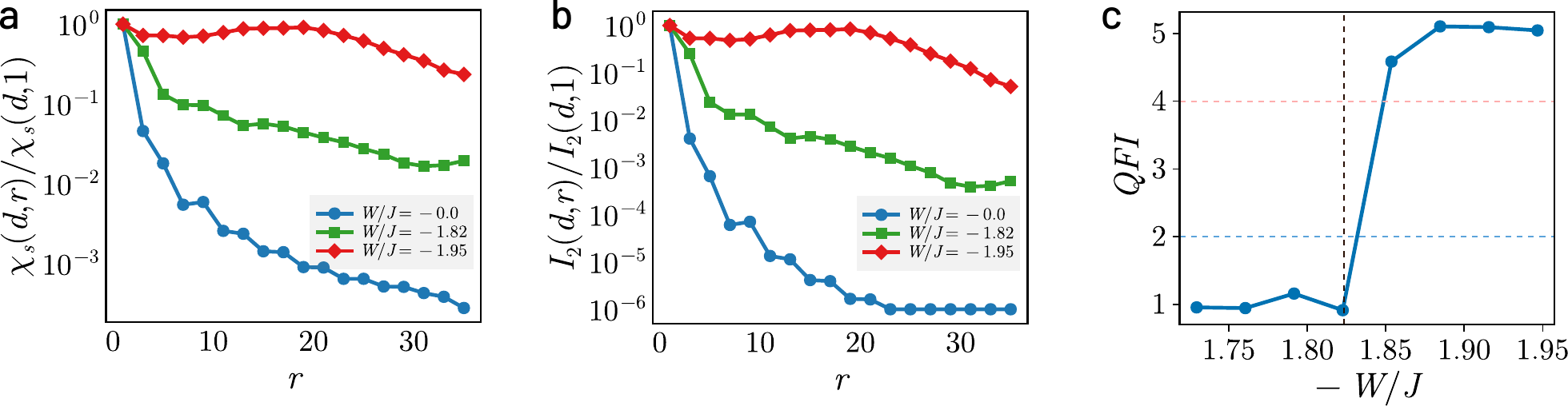}
    \caption{Spin-flip correlation $\braket{{\bf S}_d \cdot {\bf S}_r}$ (a) and mutual information $I_{2}(d,r)$ (b) between the impurity spin and conduction bath local spin density as a function of the distance {\bf r} between them, normalised against the value at $r=1$. Both decay very quickly in the FL phase (blue), but show long-ranged behaviour in the NFL phase (green and red), extending to the edges of the system at the critical point (red). (c): Evolution of the Quantum Fisher Information $F_Q$ for a nearest-neighbour spin-flip operator $\mathcal{O} = \sum_{i \in \text{odd}}(S_i^+S_{i+1}^- + \text{h.c.})$ through the first Lifshitz transition and the pseudogap. The vertical dashed line marks the onset of the PG. To the left of it, the QFI in the Fermi liquid phase shows at most bipartite entanglement ($F_Q < 2$ (below blue dashed line)), while the PG shows the presence of multipartite entanglement upto 5 parties ($F_Q > 4$ (above red dashed line)).}
    \label{longranged}
\end{figure*}

\noindent\textbf{\large Non-local nature of the Pseudogap}\\
In Fig.~\ref{longranged}(a) and (b), the spin-flip correlations and mutual information between the impurity spin and conduction bath sites respectively are observed to undergo a crossover within the PG, from a short-ranged behaviour at its onset, to a long-ranged behaviour as the Mott transition approaches. The entanglement is also observed to be multipartite in nature: in Fig.~\ref{longranged}(c), the quantum Fisher information (QFI)~\cite{Hauke2016} computed for the ground state wavefunction using an operator corresponding to the sum of local spin-flip exchange processes shows a jump at the onset of the PG. Further, the FL is observed to possess bi-partite entanglement while the NFL of the PG phase displays pentapartite entanglement~\cite{balut2025,mazza2024}.

\noindent
These striking results imply that the Mott transition observed by us lies beyond the local quantum criticality scenario~\cite{Si2001}. Instead,  we observe the PG phase to be a novel state of strongly interacting quantum matter emergent from the breakdown of local Kondo screening. This state is described by a quantum critical Fermi surface with NFL Fermi arcs that display increasingly critical behaviour, i.e., dynamics described by non-local quantum fluctuations, and excitations that become truly long-ranged close to the transition.

\vspace{0.25cm}
\par\noindent\textbf{\large Exactly solvable nodal Non-Fermi liquid at Mott Criticality}\\
Very close to the transition, the excitations of the nodal NFL correspond to those of a Hatsugai-Kohmoto model~\cite{Baskaran1991,Hatsugai1992}. This insight is obtained from a perturbation-theoretic treatment of the RG fixed point Hamiltonian of the impurity model for $W/J\lesssim (W/J)_{\text{PG}}$, by considering the effects of a small fixed point Kondo scattering probability \(J^*\) in the backdrop of a larger bath interaction parameter \(|W|\). This yields the HK model~\cite{Baskaran1991,Hatsugai1992} as the singular part of the effective Hamiltonian arising from forward scattering processes (see Sec.4 of \cite{suppmat}):
\begin{equation}\begin{aligned}\label{HKModel}
	\Delta \tilde H_{{\bf q}_1 = {\bf q}_2} = \sum_{{\bf q},\sigma}\epsilon_{{\bf q}}{n}_{{\bf q},\sigma} + u\sum_{{\bf q}, \sigma}n_{{\bf q} \sigma} n_{{\bf q} \bar\sigma}~,
\end{aligned}\end{equation}
where the number operator \(n_{{\bf q} \sigma} = \phi^\dagger_{{\bf q}, \sigma} \phi_{{\bf q}, \sigma}\) pertains to emergent fermionic relative modes $\phi_{{\bf q}, \sigma} = \frac{1}{\sqrt 2}\left(c_{{\bf N}_1 + {\bf q},\sigma} - c_{{\bf N}_1 + {\bf Q}_1 - {\bf q}, \sigma}\right)$ that are shifted by an excitation momentum \({\bf q}\) away from the nodal point \({\bf N}_1 = \left(\pi/2, \pi/2\right)\) and its 
partner \({\bf N}_1 + {\bf Q}_1 = \left(-\pi/2, -\pi/2\right)\). The kinetic energy \(\epsilon_{\bf q}\) and interaction energy \(u\sim J^{2}/W\) are dispersion and Kondo scales renormalised by conduction bath correlations (see Sec.4 of \cite{suppmat}).

\noindent
Consequently, the resulting NFL metal of the lattice model involves long-lived excitations of multiple \({\bf k}\)-states, and manifest in the form of a divergent one-particle self-energy at the non-interacting FS~\cite{Phillips2020}:
$\Sigma_{{\bf q}}(\omega) = -u^2/4(\omega - \epsilon_{\bf q})$~, such that $\Sigma_{\epsilon_{\bf q}=0}(\omega \to 0)  \to \infty$. This zero-frequency self-energy pole presages the transition into a Mott insulating phase, where it marks a hard gap in the spectral function for charge excitations. This is consistent with our findings for the lattice (Fig.~\ref{chargeCorr}(c)) and impurity self-energies~(Fig.\ref{selfEnergy} (c)). The nodal Mott metal thus comprises of a Greens function zero at the Fermi energy together with an anistropic massless Dirac dispersion, leading to a non-zero Chern number~\cite{vafekvishwanath2014,morimoto2016,calderon2025}. This topological index survives into the Mott insulator.

\noindent For small but non-zero values of \(\omega - \epsilon_{\bf k}\), we obtain a quasiparticle residue that vanishes with $\omega$, \(Z_\text{imp} \sim \omega^2/U^2 \). The scattering rate of this singular NFL possesses a sharp peak at the FS (\(\omega=\epsilon_{\bf k_{\mathrm{F}}}\)): \(\Gamma \sim U^2\delta(\omega - \epsilon_{\bf k})\), consistent with the sharply peaked Lorentzian $\Sigma''_{{\bf k}}(\omega)\sim \omega^{-2}$ captured in Fig.\ref{selfEnergy}(a). This quantum critical NFL metal is an example of a strongly coupled scale-invariant form of quantum matter. The exact solution for eq.\eqref{HKModel} reveals the presence of low-energy excitations comprised of holons and doublons~\cite{Hatsugai1992}. These features point to the nodal NFL as a long-ranged and multipartite entangled, strongly interacting scale-invariant state of quantum matter 
~\cite{Georgi2007PRL,Georgi2007,Phillips2013Unparticles,PhillipsLectures2014} that are completely disconnected from the quasiparticles of the FL. Following the arguments laid out in~\cite{phillips2022}, at finite temperatures, such a scale-invariant highly entangled non-Fermi liquid is likely associated with Planckian dissipation (i.e., $\Sigma'' (T)\sim k_{B}T$) and a resistivity that varies linearly with temperature ($\rho\sim T$)~\cite{zaanen2019,legros2019}. Additionally, we observe that the nodal metal possesses pairing fluctuations (see Sec.4 of \cite{suppmat}) that can become dominant upon doping~\cite{Phillips2020}.
\vspace{0.25cm}
\par\noindent\textbf{\large Pseudogap as a strongly coupled phase of quantum matter}\\
We now unveil an organising principle that leads to the remarkable properties observed above for the strongly interacting NFL of the PG phase. The existence of a sharp connected FS at $T=0$ can be understood as the existence of a topologically protected manifold of gapless chiral excitations in ${\bf k}$-space at the FS~\cite{Heath_2020}. The FS is characterised by a topological index corresponding to an anomaly in the quantum many-body theory for electrons, and can be understood as a generalised symmetry of such a system~\cite{lanave2025,McGreevy2023}. A theorem by Luttinger and Ward~\cite{luttinger1960ground} shows that a count of the physical charge (known as Luttinger's volume) is identical to the topological index (a so-called homotopy charge known as Luttinger's count) even in the presence of electronic interactions that do not disturb the FS. We will now argue that the emergence of antinodal Luttinger surfaces involve a disconnection of the FS (into Fermi arcs) and that, by following La Nave et al.~\cite{lanave2025}, the accompanying change in its topological properties leads to the existence of gapless NFL excitations that are non-local in nature.

\noindent
The antinodal Luttinger surfaces arise from the splitting of double poles of the single-particle Greens function on the FS into poles lying on opposite complex half-planes, together with zeros that are pinned at the FS. These changes in the analytic structure of the single-particle Greens function have important consequences. First, the emergent zeros break a $\mathbb{Z}_{2}$ symmetry of the FS~\cite{Anderson2001,Huang2022,lanave2025}. This symmetry is guaranteed within FL theory because the presence of quasiparticles vanishingly close to the FS allows the interchange of spin and charge degrees of freedom, promoting the separate $SU(2)$ symmetries of spin and charge to the larger symmetry of $O(4) = SU(2) \times SU(2) \times \mathbb{Z}_2$~\cite{Anderson2001}. The insertion of zeros of the Greens function at the Fermi surface in the PG, and the associated splitting of the Greens function, breaks this $\mathbb{Z}_{2}$ symmetry by placing the pole for the spin excitation in one half of the complex plane while placing that for the charge excitation in the other half~\cite{su2025anomalies,Huang2022}.

\noindent
Second, they signal a divergent electronic self-energy as a function of the wavevector ${\bf k}$, render ill-behaved the Luttinger-Ward functional of the interacting electronic problem, and violate the generalised symmetry encoded within it. The changes in the pole structure change the Luttinger count topological invariant, while the zeros give rise to an additional topological contribution (linked to the Adler-Bell-Jackiw-type chiral anomaly)~\cite{adler1969axial,bell1969pcac,Altshuler_1998,calderon2025}. As a consequence of the half-filled particle-hole symmetric nature of the system at hand, Luttinger's volume is preserved upon taking into account topological contributions from both the Luttinger count {\it and} the zeros~\cite{seki2017topological}. Importantly, La Nave et al.~\cite{lanave2025} argue, following recent developments in understanding generalised symmetries~\cite{Casini2021Symmetries}, that the additional anomaly arising from the Luttinger surfaces guarantees the existence of gapless NFL excitations that are non-local in nature.  

\noindent
Thus, the drastic change in nature of the real-space excitations - from locally well-defined Landau quasiparticles of the FL to the increasingly nonlocal excitations of the NFL Fermi arcs in the PG phase - appears to be dictated by a topological principle and, therefore, robust under renormalisation. This signals the NFL Fermi arcs of the PG as an emergent phase of strongly interacting quantum matter - which we dub the {\it Mott metal} - that is the parent metal of the MI. The same topological principle connects the nonlocal unparticle-like gapless excitations~\cite{Georgi2007PRL,Georgi2007,Phillips2013Unparticles,PhillipsLectures2014} of the scale-invariant nodal NFL of the HK model observed precisely at the Mott critical point to those of the rest of PG phase, e.g., a universal scaling of $\Sigma''_{{\bf k}}(\omega)\sim (a+b\omega^{2})^{-1}$ of the NFL throughout the PG phase (Fig.~\ref{selfEnergy} (a)) and whose $\omega=0$ value continues to grow upon violating the MIR bound. 
\vspace{0.25cm}
\par\noindent\textbf{\large Conclusions}\\
Our analysis reveals that the Mott transition proceeds via a continuous evolution through a PG regime characterized by a singular NFL metal - the Mott metal - with deconfined holon–doublon excitations confined to nodal Fermi arcs. As the system approaches criticality, this metallic phase exhibits increasingly non-local correlations and a divergent self-energy, signaling the breakdown of Landau quasiparticles and the onset of long-range quantum entanglement. Anchored in two-channel Kondo dynamics at intermediate scales and governed by Hatsugai–Kohmoto physics near the critical endpoint, the Mott metal provides a unified framework for understanding anomalous metallicity in strongly correlated systems. Its fate under finite doping presents a compelling direction for future investigation.

\bibliography{tilingProject}

\vspace{0.25cm}
\par\noindent\textbf{Acknowledgments.}
We acknowledge National Supercomputing Mission (NSM) for providing computing resources of ‘PARAM RUDRA’ at Aruna Asaf Ali Marg, near Vasant Kunj, Vasant Kunj, New Delhi, Delhi 110067, which is implemented by C-DAC and supported by the Ministry of Electronics and Information Technology (MeitY) and Department of Science and Technology (DST), Government of India. SL thanks the SERB, Govt. of India for funding through MATRICS grant MTR/2021/000141 and Core Research Grant CRG/2021/000852. AM thanks IISER Kolkata for funding through a JRF and an SRF. A.M. would also like to acknowledge the SERB-MATRICS grant (Grant No. MTR/2022/000636) from the Science and Engineering Research Board (SERB) for funding. S.L. thanks Mayank Shreshtha for designing Fig.1.

\vspace{0.25cm}
\par\noindent\textbf{Author Contributions.}
A.M., S.R.H. and S.L. conceived and designed the research. A.M. performed the calculations and analysis with support from all the authors. A.M., S.R.H., A.M. and S.L. wrote the manuscript with inputs from all authors. All work was supervised by S. L. 

\vspace{0.25cm}
\par\noindent\textbf{Code and Data availability.}
The codes created and datasets generated during this research are available from the corresponding author upon reasonable request.

\vspace*{0.5cm}
\par\noindent\textbf{\large Models and Methods}\\
\noindent\textbf{Lattice-embedded Impurity Model}\\
We state the detailed Hamiltonian of the auxiliary model here.
\begin{equation}
\label{impurityModel}
\mathcal{H}_\text{aux} = H_\text{imp} + H_\text{coup} + H_\text{cbath},
\end{equation}
where the particle-hole symmetric impurity term is
\begin{equation}
H_\text{imp} = - \frac{U}{2} \left( \hat{n}_{d \uparrow} - \hat{n}_{d \downarrow} \right)^2,
\end{equation}
describing on-site Coulomb interactions at the impurity. Here, $c^\dagger_{d\sigma}$ creates an electron with spin $\sigma$ at the impurity site, and $\hat{n}_{d\sigma} = c^\dagger_{d\sigma} c_{d\sigma}$ denotes the corresponding number operator.

\noindent
The impurity couples to four nearest-neighbour bath sites $c_{Z\sigma}$ through both hybridization and Kondo exchange:
\begin{equation}\begin{aligned}
H_\text{coup} = \frac{J}{2} \hspace{-0.25cm}\sum_{\sigma_1,\sigma_2,Z} 
%\sum_{Z} 
{\bf S}_d \cdot c^\dagger_{Z\sigma_1} {\boldsymbol \tau}_{\sigma_1 \sigma_2} c_{Z\sigma_2} 
- V \sum_{\sigma,Z} \left( c^\dagger_{Z\sigma} c_{d\sigma} + \text{h.c.} \right),    
\end{aligned}\end{equation}
where ${\boldsymbol \tau}$ are the Pauli matrices, and $Z$ indexes the four neighbouring bath sites. The half-filled bath includes nearest neighbour hopping and on-site correlations on the site neighbouring the impurity:
\begin{equation}
H_\text{cbath} = \sum_{\bf k} \epsilon_{\bf k} c^\dagger_{{\bf k},\sigma} c_{{\bf k},\sigma} - \frac{W}{2} \sum_{Z} \left( \hat{n}_{Z\uparrow} - \hat{n}_{Z\downarrow} \right)^2,
\end{equation}
where $\epsilon_{\bf k} = -2t(\cos k_x + \cos k_y)$, and $W$ parametrizes local repulsion in the bath. This frustrates Kondo screening and drives the system towards local-moment formation. A crucial feature of this construction is that the Kondo coupling acquires a momentum structure upon Fourier transforming:
\begin{equation}
J_{{\bf k}, {\bf k}^\prime} = \frac{J}{2} \left[ \cos(k_x - k^\prime_x) + \cos(k_y - k^\prime_y) \right],
\end{equation}
which respects the $C_4$ symmetry of the square lattice.
\vspace{0.25cm}
\par\noindent\textbf{Tiling the impurity model}\\
\noindent
We now provide details on how the effective lattice Hamiltonian is constructed from the underlying impurity model. As described in the main text, the auxiliary model consists of a single impurity embedded in a correlated bath:

\begin{equation}
H(\mathbf{r}_d) = H_{\text{imp}} + H_{\text{cbath}} + H_{\text{imp-cbath}},
\end{equation}
where the impurity Hamiltonian is
\begin{equation}
H_{\text{imp}} = -\frac{U}{2} \left( \hat{n}_{\mathbf{r}_d\uparrow} - \hat{n}_{\mathbf{r}_d\downarrow} \right)^2 - \eta \sum_\sigma \hat{n}_{\mathbf{r}_d\sigma},
\end{equation}
and the conduction bath is described by
\begin{equation}
\begin{aligned}
H_{\text{cbath}} = & -\frac{1}{\sqrt{\mathcal{Z}}} t \sum_{\langle \mathbf{r}_i, \mathbf{r}_j \rangle \ne \mathbf{r}_d; \sigma} \left( c^\dagger_{\mathbf{r}_i\sigma} c_{\mathbf{r}_j\sigma} + \text{h.c.} \right) \\
& - \frac{1}{2\mathcal{Z}} W \sum_{\mathbf{z} \in \text{NN}(\mathbf{r}_d)} \left( \hat{n}_{\mathbf{z}\uparrow} - \hat{n}_{\mathbf{z}\downarrow} \right)^2 - \mu \sum_{\mathbf{r}_i \ne \mathbf{r}_d, \sigma} \hat{n}_{\mathbf{r}_i\sigma},
\end{aligned}
\end{equation}
and the impurity–bath hybridization term is
\begin{equation}
\begin{aligned}
H_{\text{imp-cbath}} = & \frac{J}{\mathcal{Z}} \sum_{\sigma, \sigma'} \sum_{\mathbf{z} \in \text{NN}(\mathbf{r}_d)} \mathbf{S}_{\mathbf{r}_d} \cdot \boldsymbol{\tau}_{\sigma \sigma'} c^\dagger_{\mathbf{z}\sigma} c_{\mathbf{z}\sigma'} \\
& - \frac{V}{\sqrt{\mathcal{Z}}} \sum_{\sigma} \sum_{\mathbf{z} \in \text{NN}(\mathbf{r}_d)} \left( c^\dagger_{\mathbf{r}_d\sigma} c_{\mathbf{z}\sigma} + \text{h.c.} \right),
\end{aligned}
\end{equation}
where $\boldsymbol{\tau} = (\tau_x, \tau_y, \tau_z)$ are Pauli matrices, and ${\bf r}_d$ denotes the impurity position.

\noindent
The tiled lattice Hamiltonian is generated by symmetrically translating this impurity model:
\begin{equation}
\mathcal{H}_{\text{tiled}} = \mathcal{T}[H(\mathbf{r}_d)] = \sum_{\mathbf{r}} T^\dagger(\mathbf{r} - \mathbf{r}_d) H(\mathbf{r}_d) T(\mathbf{r} - \mathbf{r}_d),
\end{equation}
where $T(\mathbf{a})$ are many-body translation operators, defined by
\begin{equation}
T^\dagger(\mathbf{a}) \mathcal{O}(\{\mathbf{r}_i\}) T(\mathbf{a}) = \mathcal{O}(\{\mathbf{r}_i + \mathbf{a}\}).
\end{equation}
Applying this to each term yields:
\begin{equation}\hspace{-0.5cm}\begin{aligned}
\mathcal{T}[H_{\text{cbath}}] &= -\frac{(N - 2)t}{\sqrt{\mathcal{Z}}} \sum_{\langle \mathbf{r}_i, \mathbf{r}_j \rangle; \sigma} \left( c^\dagger_{\mathbf{r}_i\sigma} c_{\mathbf{r}_j\sigma} + \text{h.c.} \right) \\
&\quad - \frac{1}{2} W \sum_{\mathbf{r}} \left( \hat{n}_{\mathbf{r}\uparrow} - \hat{n}_{\mathbf{r}\downarrow} \right)^2 - \mu (N - 1) \sum_{\mathbf{r}, \sigma} \hat{n}_{\mathbf{r}\sigma}, \\
\mathcal{T}[H_{\text{imp}}] &= -\frac{U}{2} \sum_{\mathbf{r}} \left( \hat{n}_{\mathbf{r}\uparrow} - \hat{n}_{\mathbf{r}\downarrow} \right)^2 - \eta \sum_{\mathbf{r}, \sigma} \hat{n}_{\mathbf{r}\sigma}, \\
\mathcal{T}[H_{\text{imp-cbath}}] &= \sum_{\langle \mathbf{r}_i, \mathbf{r}_j \rangle} \left[ \frac{2J}{\mathcal{Z}} \mathbf{S}_{\mathbf{r}_i} \cdot \mathbf{S}_{\mathbf{r}_j} - \frac{2V}{\sqrt{\mathcal{Z}}} \sum_\sigma \left( c^\dagger_{\mathbf{r}_i\sigma} c_{\mathbf{r}_j\sigma} + \text{h.c.} \right) \right],
\end{aligned}\end{equation}
with local spin operators defined as
\begin{equation}
\mathbf{S}_{\mathbf{r}} = \sum_{\sigma, \sigma'} c^\dagger_{\mathbf{r}\sigma} \boldsymbol{\tau}_{\sigma \sigma'} c_{\mathbf{r}\sigma'}.
\end{equation}

\noindent
To avoid overcounting of the non-interacting bath terms, we subtract extra copies:
\begin{equation}
\mathcal{H}_{\text{cbath-nint}} = -\frac{1}{\sqrt{\mathcal{Z}}} t \sum_{\langle \mathbf{r}_i, \mathbf{r}_j \rangle; \sigma} \left( c^\dagger_{\mathbf{r}_i\sigma} c_{\mathbf{r}_j\sigma} + \text{h.c.} \right) - \mu \sum_{\mathbf{r}, \sigma} \hat{n}_{\mathbf{r}\sigma}.
\end{equation}
The final reconstructed Hubbard–Heisenberg model then reads:
\begin{equation}
\begin{aligned}
\mathcal{H}_{\text{HH}} = & -\frac{1}{\sqrt{\mathcal{Z}}} \tilde{t} \sum_{\langle \mathbf{r}_i, \mathbf{r}_j \rangle; \sigma} \left( c^\dagger_{\mathbf{r}_i\sigma} c_{\mathbf{r}_j\sigma} + \text{h.c.} \right) - \tilde{\mu} \sum_{\mathbf{r}, \sigma} \hat{n}_{\mathbf{r}\sigma} \\
& + \frac{\tilde{J}}{\mathcal{Z}} \sum_{\langle \mathbf{r}_i, \mathbf{r}_j \rangle} \mathbf{S}_{\mathbf{r}_i} \cdot \mathbf{S}_{\mathbf{r}_j} - \frac{1}{2} \tilde{U} \sum_{\mathbf{r}} \left( \hat{n}_{\mathbf{r}\uparrow} - \hat{n}_{\mathbf{r}\downarrow} \right)^2,
\end{aligned}
\end{equation}
where the effective lattice couplings are given by:
\begin{equation}
\tilde{t} = t + 2V,\quad \tilde{U} = U + W,\quad \tilde{\mu} = 2\mu + \eta,\quad \tilde{J} = 2J.
\end{equation}
In summary, the tiling prescription reconstructs the full Hamiltonian as
\begin{equation}
\mathcal{H}_{\text{tiled}} = \sum_{\mathbf{r}} H(\mathbf{r}) - N \mathcal{H}_{\text{cbath-nint}},
\end{equation}
effectively embedding the impurity physics into a global lattice model.

\newpage
\onecolumngrid
\section*{Supplementary Information}
\section{Tiling basics}
\subsection{Translation symmetry and a conserved total momentum}
The tiled Hamiltonian is symmetric under global many-body translations 
%of the kind defined in eq.~\ref{translationDefinition}, 
by arbitrary lattice spacings:
\begin{equation}\begin{aligned}\label{translationSymmetry}
	&T({\bf a})^\dagger \mathcal{H}_\text{tiled} T({\bf a}) = T({\bf a})^\dagger\sum_{{\bf r}}\mathcal{H}_\text{aux}({\bf r})T({\bf a}) = \sum_{{\bf r}}\mathcal{H}_\text{aux}({\bf r + a}) = \sum_{{\bf r}^\prime}\mathcal{H}_\text{aux}({\bf r}^\prime)\\
	&T({\bf a})^\dagger\sum_{{\bf r}}\mathcal{H}_\text{cbath-nint}T({\bf a}) = \mathcal{H}_\text{cbath-nint}\\
	&\implies T({\bf a})^\dagger\mathcal{H}_\text{tiled}T({\bf a}) = \mathcal{H}_\text{tiled}~.
\end{aligned}\end{equation}
In the first equation, we used the fact that the translation operator simply translates the auxiliary model at the position \({\bf r}\) into another one at the position \({\bf r} + {\bf a}\). Since both are part of the summation, the summation remains unchanged. The second equation uses the fact that the Hamiltonian \(\mathcal{H}_\text{cbath-nint}\) is that of a tight-binding model and is therefore translation-invariant. The fact that the Hamiltonian \(\mathcal{H}_\text{tiled}\) commutes with the many-body translation operator implies that the total crystal momentum \(\vec k\) is a conserved quantity.

\subsection{Form of the eigenstates: Bloch's theorem}
In the tight-binding approach to lattice problems, the full Hamiltonian is described by adding the localised Hamiltonians at each site, and the full eigenstate \(\ket{\Psi}\) is then obtained by constructing liner combinations of the eigenstates \(\ket{\psi_i}\) of the local Hamiltonians such that \(\ket{\Psi}\) satisfies Bloch's theorem: \(\ket{\Psi_{\bf k}} = \sum_{i} e^{i {\bf k}\cdot{\bf r}_i} \ket{\psi_i}\), where \({\bf r}_i\) sums over the positions of the local Hamiltonians. Bloch's theorem ensures that eigenstates satisfy the following relation under a translation operation by an arbitrary number of lattice spacings \({n\bf a}\):
\begin{equation}\begin{aligned}
	T^\dagger(n{\bf a})\ket{\Psi_{{\bf k}}} = \sum_{i} e^{i {\bf k}\cdot{\bf r}_i} \ket{\psi_{i + n}} = e^{-in{\bf k}\cdot{\bf a}}\ket{\Psi_{\bf k}}
\end{aligned}\end{equation}
%The definition and some properties of these global translation operations were provided in Appendix~\ref{BlochProperties}. It was shown there that they share eigenstates with the total momentum operator. 
In a lattice model, the continuous translation symmetry is lowered to its discrete form: the total {\it crystal} momentum is conserved by any scattering process. As a result, the eigenstates can be labelled using the combined index \(s = \left({\bf k}, n\right) \) where \({\bf k}\) is the total crystal momentum and \(n\) is a band index \(n\).

The eigenstates \(\ket{\Psi_{s}}\) (\(s = \left({\bf k}, n\right)\)) of the lattice Hamiltonians 
%obtained using eq.~\ref{tilingPrescriptionFinal} also 
enjoy a {\it many-body} Bloch's theorem~[53]
%\cite{stoyanova}, 
because the tiling procedure restores the translation symmetry of the Hamiltonian (as shown in eq.~\ref{translationSymmetry}). This means that the {\it local} eigenstates \(\ket{\psi_n\left({\bf r}_d\right)}\) (with the impurity located at an arbitrary position \({\bf r}_d\)) of the unit cell auxiliary model Hamiltonian \(\mathcal{H}_\text{aux}({\bf r}_d)\) %defined in eq.~\ref{unitCellHamiltonian} 
can be used to construct eigenstates of the lattice Hamiltonian. The index \(n(=0,1,\ldots)\) in the subscript indicates that it is the \(n^\text{th}\) eigenstate of the auxiliary model.

The state \(\ket{\psi_n\left({\bf r}_d\right)}\) does not specify the position of the zeroth site, because the unit cell Hamiltonian \(\mathcal{H}_\text{aux}({\bf r}_d)\) itself has been averaged over \(\mathcal{Z}\) zeroth sites. Accordingly, we can express the averaged eigenstate \(\ket{\psi_n\left({\bf r}_d\right)}\) as
\begin{equation}\begin{aligned}
	\ket{\psi_n\left({\bf r}_d\right)} = \frac{1}{\sqrt\mathcal{Z}}\sum_{{\bf z} \in \text{NN}({\bf r}_d)}\ket{\psi_n\left({\bf r}_d, {\bf z}\right)}~,
\end{aligned}\end{equation}
where \(\ket{\psi_n\left({\bf r}_d, {\bf z}\right)}\) is an auxiliary model eigenstate with the impurity and zeroth sites placed at \({\bf r}_d\) and \({\bf z}\). With this in mind, the following unnormalised combination of the auxiliary model eigenstates satisfies a many-particle equivalent of Bloch's theorem~[53]:
%\cite{stoyanova}:
\begin{equation}\begin{aligned}\label{eigenstateProposal}
	\ket{\Psi_{s}} \equiv \ket{\Psi_{{\bf k}, n}} &= \frac{1}{\sqrt N}\sum_{{\bf r}_d} e^{i {\bf k}\cdot{\bf r}_d} \ket{\psi_{n}\left({\bf r}_d\right)} = \frac{1}{\sqrt{\mathcal{Z} N}}\sum_{{\bf r}_d}\sum_{{\bf z} \in \text{NN}({\bf r}_d)} e^{i {\bf k}\cdot{\bf r}_d} \ket{\psi_{n}\left({\bf r}_d, {\bf z}\right)}~,
\end{aligned}\end{equation}
where \(N\) is the total number of lattice sites and \({\bf r}_d\) is summed over all lattice spacings. The set of \(n=0\) states form the lowest band in the spectrum of the lattice, while higher values of \(n\) produce the more energetic bands. The ground state \(s = s_0\) is obtained by setting \({\bf k}\) and \(n\) to 0:
\begin{equation}\begin{aligned}\label{groundstateProposal}
	\ket{\Psi_\text{gs}} \equiv \ket{\Psi_{s_0}} &= \frac{1}{\sqrt N}\sum_{{\bf r}_d} e^{i {\bf k}\cdot{\bf r}_d} \ket{\psi_\text{gs}\left({\bf r}_d\right)} = \frac{1}{\sqrt{\mathcal{Z} N}}\sum_{{\bf r}_d}\sum_{{\bf z} \in \text{NN}({\bf r}_d)} e^{i {\bf k}\cdot{\bf r}_d} \ket{\psi_\text{gs}\left({\bf r}_d, {\bf z}\right)}
\end{aligned}\end{equation}

\section{Derivation of unitary RG equations for the lattice-embedded impurity model}

\subsection{The unitary renormalisation group method}
In order to obtain the various low-energy phases of our impurity model, we perform a scaling analysis of the associated Hamiltonian using the recently developed unitary renormalisation group (URG) method ~[19,39,40].
%\cite{anirbanurg1,anirbanurg2}. 
The method has been applied successfully on a wide variety of problems of correlated fermions~\cite{santanukagome,1dhubjhep1,anirbanmott1,siddharthacpi,anirban_kondo,Patra_2023}. The method proceeds by resolving quantum fluctuations in high-energy degrees of freedom, leading to a low-energy Hamiltonian with renormalised couplings and new emergent degrees of freedom. Typically, for a system with Fermi energy \(\epsilon_F\) and bandwidth \(E_N\), the sequence of isoenergetic shells \(\left\{E_{(j)}\right\}, E_{j}\in \left[E_0, E_N\right] \) define the states whose quantum fluctuations we sequentially resolve. The momentum states lying on shells \(E_N\) that are far away from the Fermi surface comprise the UV states, while those on shells near the Fermi surface comprise the IR states.

As a result of the URG transformations, the Hamiltonian \(H_{(j)}\) at a given RG step \(j\) involves scattering processes between the \(k-\)states that have energies lower than \(D_{(j+1)}\). The unitary transformation \(U_{(j)}\) is then defined so as to remove the number fluctuations of the currently most energetic set of states \(D_{(j)}\)~[39,40]:
%\cite{anirbanurg1,anirbanurg2}:
\begin{eqnarray}
	H_{(j-1)} = U_{(j)} H_{(j)} U^\dagger_{(j)}~, \text{such that} ~\left[H_{(j-1)}, \hat n_{j}\right] =0~.
\end{eqnarray}
The eigenvalue of $\hat{n}_{j}$ has, thus, been rendered an integral of motion (IOM) under the RG transformation.

The unitary transformations can be expressed in terms of a generator \(\eta_{(j)}\) that has fermionic algebra~[39,40]:
%\cite{anirbanurg1,anirbanurg2}:
\begin{eqnarray}
	\label{unitary}
	U_{(j)} = \frac{1}{\sqrt 2}\left(1 + \eta_{(j)} - \eta_{(j)}^\dagger\right)~,~ \quad\left\{ \eta_{(j)},\eta_{(j)}^\dagger \right\} = 1~,
\end{eqnarray}
where \(\left\{\cdot\right\}\) is the anticommutator. The unitary operator \(U_{(j)}\) that appears in Eq.~\eqref{unitary} can be cast into the well-known general form \(U = e^\mathcal{S}, \mathcal{S} = \frac{\pi}{4}\left( \eta^\dagger_{(j)} - \eta_{(j)} \right)\) that a unitary operator can take, defined by an anti-Hermitian operator \(\mathcal{S}\). The generator \(\eta_{(j)}\) is given by the expression~[39,40]:
%\cite{anirbanurg1,anirbanurg2}
\begin{eqnarray}
	\eta^\dagger_{(j)} = \frac{1}{\hat \omega_{(j)} - \text{Tr}\left(H_{(j)} \hat n_{j}\right) } c^\dagger_{j} \text{Tr}\left(H_{(j)}c_{j}\right)~.
\end{eqnarray}
The operators \(\eta_{(j)},\eta^\dagger_{(j)}\) behave as the many-particle analogues of the single-particle field operators \(c_j,c^\dagger_j\) - they change the occupation number of the single-particle Fock space \(\ket{n_j}\).  The important operator \(\hat \omega_{(j)}\) originates from the quantum fluctuations that exist in the problem because of the non-commutation of the kinetic energy terms and the interaction terms in the Hamiltonian:
\begin{eqnarray}
	\hat \omega_{(j)} = H_{(j-1)} - H^i_{(j)}~.
	\label{omega}
\end{eqnarray}
\(H^i_{(j)}\) is the part of \(H_{(j)}\) that commutes with \(\hat n_j\) but does {\it not} commute with at least one \(\hat n_l\) for \(l < j\). The RG flow continues up to energy \(D^*\), where a fixed point is reached from the vanishing of the RG function. 
Detailed comparisons of the URG with other methods (e.g., the functional RG, spectrum bifurcation RG etc.) can be found in Refs.~[39,40].
%\cite{anirbanmott1,anirbanurg1}.

\subsection{RG scheme}
At any given step \(j\) of the RG procedure, we decouple the states \(\left\{ {\bf q} \right\} \) on the isoenergetic surface of energy \(\varepsilon_j\). The diagonal Hamiltonian \(H_D\) for this step consists of all terms that do not change the occupancy of the states \(\left\{{\bf q}\right\}\):
\begin{equation}\begin{aligned}
	H_D^{(j)} = \varepsilon_j\sum_{q,\sigma}\tau_{q,\sigma} + \frac{1}{2}\sum_{{\bf q}}J_{{\bf q}, {\bf q}}S_d^z\left(\hat n_{{\bf q}, \uparrow} - \hat n_{{\bf q}, \downarrow}\right) - \frac{1}{2}\sum_{{\bf q}}W_{\bf q}\left(\hat n_{{\bf q}, \uparrow} - \hat n_{{\bf q}, \downarrow}\right)^2~,
\end{aligned}\end{equation}
where \(\tau = \hat n - 1/2\) and \(W_{{\bf q}}\) is a shorthand for \(W_{{\bf q},{\bf q},{\bf q},{\bf q}}\). The three terms, respectively, are the kinetic energy of the momentum states on the isoenergetic shell that we are decoupling, the spin-correlation energy between the impurity spin and the spins formed by these momentum states and, finally, the local correlation energy associated with these states arising from the \(W\) term. The off-diagonal part of the Hamiltonian on the other hand leads to scattering in the states \(\left\{ {\bf q} \right\} \). We now list these terms, classified by the coupling they originate from.

%\subsubsection*{Arising from the Kondo spin-exchange term}
\vspace{0.25cm}
\par\noindent\textbf{Arising from the Kondo spin-exchange term}\\
\begin{equation}\begin{aligned}
	T_{KZ1}^\dagger + T_{KZ1} &= \frac{1}{2}\sum_{{\bf k}, {\bf q}, \sigma}\sigma J_{{\bf k}, {\bf q}} S_d^z  \left[c^\dagger_{{\bf q}\sigma}c_{{\bf k},\sigma} + \text{h.c.}\right],\\
	T_{KZ2}^\dagger + T_{KZ2} &= \frac{1}{2}\sum_{{\bf q}, \sigma}\sigma J_{{\bf q}, {\bf \bar q}} S_d^z  \left[c^\dagger_{{\bf q}\sigma}c_{{\bf \bar q},\sigma} + \text{h.c.}\right],\\
	T_{KT1}^\dagger + T_{KT1} &= \frac{1}{2}\sum_{{\bf k}, {\bf q}}J_{{\bf k}, {\bf q}} \left[S_d^+\left(c^\dagger_{{\bf q}\downarrow}c_{{\bf k}\uparrow} + c^\dagger_{{\bf k}\downarrow}c_{{\bf q}\uparrow}\right) + \text{h.c.}\right],\\
	T_{KT2}^\dagger + T_{KT2} &= \frac{1}{2}\sum_{{\bf q}}J_{{\bf q}, {\bf \bar q}} \left[S_d^+\left(c^\dagger_{{\bf q}\downarrow}c_{{\bf \bar q}\uparrow} + c^\dagger_{{\bf \bar q}\downarrow}c_{{\bf q}\uparrow}\right) + \text{h.c.}\right],\\
\end{aligned}\end{equation}
\vspace{0.25cm}
\par\noindent\textbf{Arising from spin-preserving scattering within conduction bath}\\
%\subsubsection*{Arising from spin-preserving scattering within conduction bath}
\begin{equation}\begin{aligned}
	T_{P1}^\dagger + T_{P1} &= -\sum_{{\bf q} \in \varepsilon_j}\sum_{{\bf k}_2, {\bf k}_3, {\bf k}_4 < \varepsilon_j}\sum_{\sigma} \left[W_{{\bf q},{\bf k}_2,{\bf k}_3,{\bf k}_4}c^\dagger_{{\bf q},\sigma}c_{{\bf k}_2,\sigma}c^\dagger_{{\bf k}_3,\sigma}c_{{\bf k}_4,\sigma} + \text{h.c.}\right]\\
	T_{P2}^\dagger + T_{P3} &= -\sum_{{\bf q} \in \varepsilon_j}\sum_{{\bf k}_2 < \varepsilon_j}\sum_{\sigma} W_{{\bf q},{\bf k}_2,{\bf \bar q}, {\bf \bar q}} c^\dagger_{{\bf q},\sigma}c_{{\bf k}_2,\sigma} n_{{\bf \bar q},\sigma} -\sum_{{\bf q} \in \varepsilon_j}\sum_{{\bf k}_1 < \varepsilon_j}\sum_{\sigma} W_{{\bf k}_1,{\bf q},{\bf q},{\bf q}} c^\dagger_{{\bf k}_1,\sigma}c_{{\bf q},\sigma} n_{{\bf q},\sigma}\\
	T_{P4} &= -\sum_{{\bf q} \in \varepsilon_j}\sum_{{\bf k}_2, {\bf k}_3 < \varepsilon_j}\sum_{\sigma}W_{{\bf q},{\bf \bar q},{\bf k}_2,{\bf k}_3} c^\dagger_{{\bf q},\sigma}c_{{\bf \bar q},\sigma}c^\dagger_{{\bf k}_2,\sigma}c_{{\bf k}_3,\sigma}\\
	T_{P5} &= -\sum_{{\bf q} \in \varepsilon_j}\sum_{{\bf k}_2, {\bf k}_3 < \varepsilon_j}\sum_{\sigma} W_{{\bf q},{\bf k}_2,{\bf k}_3,{\bf \bar q}} c^\dagger_{{\bf q},\sigma}c_{{\bf k}_2,\sigma}c^\dagger_{{\bf k}_3,\sigma}c_{{\bf \bar q},\sigma}\\
		   &= +\sum_{{\bf q} \in \varepsilon_j}\sum_{{\bf k}_2, {\bf k}_3 < \varepsilon_j}\sum_{\sigma} W_{{\bf q},{\bf k}_3,{\bf k}_2,{\bf \bar q}} c^\dagger_{{\bf q},\sigma}c_{{\bf \bar q},\sigma}c^\dagger_{{\bf k}_2,\sigma}c_{{\bf k}_3,\sigma}\\
		   &= -T_{P4}
\end{aligned}\end{equation}
\vspace{0.25cm}
\par\noindent\textbf{Arising from spin-flip scattering within conduction bath}\\
%\subsubsection*{Arising from spin-flip scattering within conduction bath}
\begin{equation}\begin{aligned}
	T_{F1}^\dagger + T_{F1} &= \sum_{{\bf q} \in \varepsilon_j}\sum_{{\bf k}_2, {\bf k}_3, {\bf k}_4 < \varepsilon_j}\sum_{\sigma} \left[W_{{\bf q},{\bf k}_2,{\bf k}_3,{\bf k}_4}c^\dagger_{{\bf q},\sigma}c_{{\bf k}_2,\sigma}c^\dagger_{{\bf k}_3,\bar\sigma}c_{{\bf k}_4,\bar\sigma} + \text{h.c.}\right]\\
	T_{F2} &= \sum_{{\bf q}, {\bf q}^\prime \in \varepsilon_j}\sum_{{\bf k}_2, {\bf k}_3 < \varepsilon_j}\sum_{\sigma}W_{{\bf q},{\bf q}^\prime,{\bf k}_2,{\bf k}_3} c^\dagger_{{\bf q},\sigma}c_{{\bf q}^\prime,\sigma}c^\dagger_{{\bf k}_2,\bar\sigma}c_{{\bf k}_3,\bar\sigma}\\
	T_{F3} &= \sum_{{\bf q}, {\bf q}^\prime \in \varepsilon_j}\sum_{{\bf k}_2, {\bf k}_3 < \varepsilon_j}\sum_{\sigma} W_{{\bf q},{\bf k}_2,{\bf k}_3,{\bf q}^\prime} c^\dagger_{{\bf q},\sigma}c_{{\bf k}_2,\sigma}c^\dagger_{{\bf k}_3,\bar\sigma}c_{{\bf q}^\prime,\bar\sigma} \\
	T_{F4}^\dagger + T_{F4} &= \sum_{{\bf q}, {\bf q}^\prime, \in \varepsilon_j}\sum_{{\bf k}_1 < \varepsilon_j}\sum_{\sigma} \left[W_{{\bf q},{\bf q},{\bf q}^{\prime}, {\bf k}_1} n_{{\bf q},\sigma} c^\dagger_{{\bf q}^{\prime},\bar\sigma}c_{{\bf k}_1,\bar\sigma} + \text{h.c.}\right]
\end{aligned}\end{equation}
In all of the terms \(T_{P[i]}\) and \(T_{F[i]}\), the factor of \(1/2\) in front has been cancelled out by a factor of 2 coming from the multiple possibilities of arranging the momentum labels. We will henceforth ignore \(T_{P4}\) and \(T_{P5}\) because they cancel each other out.

The renormalisation of the Hamiltonian is constructed from the general expression
\begin{equation}\begin{aligned}
	\Delta H^{(j)} = H_X \frac{1}{\omega- H_D} H_X~.
\end{aligned}\end{equation}

The states on the isoenergetic shell \(\pm|\varepsilon_j|\) come in particle-hole pairs \(\left( {\bf q}, {\bf \bar q} \right) \) with energies of opposite signs (relative to the Fermi energy). If \({\bf q}\) is defined as the hole state (unoccupied in the absence of quantum fluctuations), it will have positive energy, while the particle state \({\bf \bar q}\) will be of negative energy and hence below the Fermi surface. To be more specific, given a state \({\bf q}\) with energy \(\pm|\varepsilon_j|\), we define its particle-hole transformed counterpart as the state \({\bf \bar q} = \boldsymbol{\pi} + {\bf q}\), having energy \(\mp|\varepsilon_j|\) and residing in the opposite quadrant of the Brillouin zone. Given this definition, we have the important property that
\begin{equation}\begin{aligned}\label{particleHoleRelation}
	J_{{\bf k}, {\bf \bar q}} &= -J_{{\bf k}, {\bf q}},\\
	W_{\left\{{\bf k}\right\}, {\bf \bar q}} &= -W_{\left\{{\bf k}\right\}, {\bf q}}~.
\end{aligned}\end{equation}

\subsection{Renormalisation of the bath correlation term {\it W}}
The bath correlation term \(W\) can undergo renormalisation only via scattering processes arising from itself. Irrespective of whether the state \({\bf q}\) being decoupled is in a particle or hole configuration in the initial many-body state, the propagator \(G = 1/(\omega - H_D)\) of the intermediate excited state is uniform, and equal to 
\begin{equation}\begin{aligned}\label{propagatorW}
	G_W = 1/\left(\omega - |\varepsilon_j|/2 + W_{\bf q}/2)\right)~,
\end{aligned}\end{equation}
where \(W_{\bf q}\) is the same whether \({\bf q}\) is above or below the Fermi surface. The \(|\varepsilon_j|/2\) in \(H_D\) arises from the excited nature of the state after the initial scattering process.
\vspace{0.25cm}
\par\noindent\textbf{Scattering arising purely from spin-preserving processes}\\
%\subsubsection*{Scattering arising purely from spin-preserving processes}
In this subsection, we calculate the renormalisation to \(W\) arising from the terms \(T_{P1}\), \(T_{P2}\) and \(T_{P3}\). The first term is
\begin{equation}\begin{aligned}
	T_{P1}^\dagger G_W T_{P3} &= \sum_\sigma\sum_{{\bf k}_1, {\bf k}_2, {\bf k}_3, {\bf k}_4} \sum_{\bf q} W_{{\bf q},{\bf k}_2,{\bf k}_3,{\bf k}_4}c^\dagger_{{\bf q},\sigma}c_{{\bf k}_2,\sigma}c^\dagger_{{\bf k}_3,\sigma}c_{{\bf k}_4,\sigma} G_W W_{{\bf k}_1,{\bf q},{\bf q},{\bf q}} c^\dagger_{{\bf k}_1,\sigma}c_{{\bf q},\sigma} n_{{\bf q},\sigma}\\
							  &= -\sum_\sigma\sum_{{\bf k}_1, {\bf k}_2, {\bf k}_3, {\bf k}_4} c^\dagger_{{\bf k}_1,\sigma} c_{{\bf k}_2,\sigma}c^\dagger_{{\bf k}_3,\sigma}c_{{\bf k}_4,\sigma} \sum_{\bf q \in \text{PS}} W_{{\bf q},{\bf k}_2,{\bf k}_3,{\bf k}_4} G_W W_{{\bf k}_1,{\bf q},{\bf q},{\bf q}}~.
\end{aligned}\end{equation}
The operators acting on the states being decoupled contract to form a number operator \(n_{{\bf q},\sigma}\) which projects the sum over \({\bf q}\) into the states that are initial occupied (particle sector, PS). 

The second such contribution is obtained by flipping the sequence of scattering processes:
\begin{equation}\begin{aligned}
	T_{P3} G_W T_{P1}^\dagger &= \sum_\sigma\sum_{{\bf k}_1, {\bf k}_2, {\bf k}_3, {\bf k}_4} \sum_{\bf q} W_{{\bf k}_1,{\bf q},{\bf q},{\bf q}} c^\dagger_{{\bf k}_1,\sigma}c_{{\bf q},\sigma} n_{{\bf q},\sigma} G_W W_{{\bf q},{\bf k}_2,{\bf k}_3,{\bf k}_4}c^\dagger_{{\bf q},\sigma}c_{{\bf k}_2,\sigma}c^\dagger_{{\bf k}_3,\sigma}c_{{\bf k}_4,\sigma} \\
							  &= \sum_\sigma\sum_{{\bf k}_1, {\bf k}_2, {\bf k}_3, {\bf k}_4} c^\dagger_{{\bf k}_1,\sigma} c_{{\bf k}_2,\sigma}c^\dagger_{{\bf k}_3,\sigma}c_{{\bf k}_4,\sigma} \sum_{\bf q \in \text{HS}} W_{{\bf q},{\bf k}_2,{\bf k}_3,{\bf k}_4} G_W W_{{\bf k}_1,{\bf q},{\bf q},{\bf q}}~.
\end{aligned}\end{equation}
By virtue of eq.~\ref{particleHoleRelation}, the product of couplings \(W_{{\bf q},{\bf k}_2,{\bf k}_3,{\bf k}_4} G_W W_{{\bf k}_1,{\bf q},{\bf q},{\bf q}}\) is the same irrespective of whether \({\bf q}\) belongs to the particle or hole sector. The two contributions therefore cancel each other. Moreover, the remaining contributions \(T_{P3}^\dagger G_W T_{P1}\) and \(T_{P1}G_W T_{P2}^\dagger\) are effectively hermitian conjugates of the two contributions considered above, and therefore also cancel each other.
\vspace{0.25cm}
\par\noindent\textbf{Scattering arising from spin-flip processes}\\
%\subsubsection*{Scattering arising from spin-flip processes}
We now come to the processes that involve spin-flips. Considering \(T_{F1}\) and \(T_{F4}\) first, we get
\begin{equation}\begin{aligned}
	T_{F1}^\dagger G_W T_{F4} &= \sum_\sigma\sum_{{\bf k}_1, {\bf k}_2, {\bf k}_3, {\bf k}_4} \sum_{\bf q} W_{{\bf q},{\bf k}_2,{\bf k}_3,{\bf k}_4}c^\dagger_{{\bf q},\sigma}c_{{\bf k}_2,\sigma}c^\dagger_{{\bf k}_3,\bar\sigma}c_{{\bf k}_4,\bar\sigma} G_W W_{{\bf k}_1, {\bf q}, {\bf q}, {\bf q}}  c^\dagger_{{\bf k}_1\sigma} c_{{\bf q}\sigma} n_{{\bf q} \bar \sigma} \\
							  &= -\sum_{1,2,3,4}\sum_\sigma c^\dagger_{{\bf k}_1\sigma} c_{{\bf k}_2\sigma} c^\dagger_{{\bf k}_3\bar\sigma} c_{{\bf k}_4\bar \sigma} \sum_{\bf q \in \text{PS}} W_{{\bf q}, {\bf k}_2, {\bf k}_4, {\bf k}_4} G_W W_{{\bf k}_1, {\bf q}, {\bf q}, {\bf q}}~,\\
	T_{F4} G_W T_{F1}^\dagger &= \sum_\sigma\sum_{{\bf k}_1, {\bf k}_2, {\bf k}_3, {\bf k}_4} \sum_{\bf q}W_{{\bf k}_1, {\bf q}, {\bf q}, {\bf q}}  c^\dagger_{{\bf k}_1\sigma} c_{{\bf q}\sigma} n_{{\bf q} \bar \sigma} G_W W_{{\bf q},{\bf k}_2,{\bf k}_3,{\bf k}_4}c^\dagger_{{\bf q},\sigma}c_{{\bf k}_2,\sigma}c^\dagger_{{\bf k}_3,\bar\sigma}c_{{\bf k}_4,\bar\sigma}  \\
							  &= \sum_{1,2,3,4}\sum_\sigma c^\dagger_{{\bf k}_1\sigma} c_{{\bf k}_2\sigma} c^\dagger_{{\bf k}_3\bar\sigma} c_{{\bf k}_4\bar \sigma} \sum_{\bf q \in \text{HS}} W_{{\bf q}, {\bf k}_2, {\bf k}_4, {\bf k}_4} G_W W_{{\bf k}_1, {\bf q}, {\bf q}, {\bf q}}~.
\end{aligned}\end{equation}
By the same arguments as in the previous subsection, these terms cancel each other out. Their hermitian conjugate contributions \(T_{F1} G_W T_{F4}^\dagger\) and \(T_{F4}^\dagger G_W T_{F1}\) also cancel out. The other two terms are \(T_{F2}\) and \(T_{F3}\), and their contributions also cancel out for the same reason:
\begin{equation}\begin{aligned}
	T_{F2} G_W T_{F2} &= \sum_\sigma\sum_{{\bf k}_1, {\bf k}_2, {\bf k}_3, {\bf k}_4} \sum_{\bf q}W_{{\bf q},{\bf \bar q},{\bf k}_3,{\bf k}_4} c^\dagger_{{\bf q},\sigma}c_{{\bf \bar q},\sigma}c^\dagger_{{\bf k}_3,\bar\sigma}c_{{\bf k}_4,\bar\sigma} G_W W_{{\bf \bar q},{\bf q},{\bf k}_1,{\bf k}_2} c^\dagger_{{\bf \bar q},\sigma}c_{{\bf q},\sigma}c^\dagger_{{\bf k}_1,\bar\sigma}c_{{\bf k}_2,\bar\sigma} \\
							  &= \sum_{1,2,3,4}\sum_\sigma c^\dagger_{{\bf k}_1\sigma} c_{{\bf k}_2\sigma} c^\dagger_{{\bf k}_3\bar\sigma} c_{{\bf k}_4\bar \sigma} \sum_{\bf q \in \text{PS}} W_{{\bf q},{\bf \bar q},{\bf k}_3,{\bf k}_4} G_W W_{{\bf \bar q},{\bf q},{\bf k}_1,{\bf k}_2}~,\\
	T_{F3} G_W T_{F3} &= \sum_\sigma\sum_{{\bf k}_1, {\bf k}_2, {\bf k}_3, {\bf k}_4} \sum_{\bf q} W_{{\bf q},{\bf k}_2,{\bf k}_3,{\bf \bar q}} c^\dagger_{{\bf q},\sigma}c_{{\bf k}_2,\sigma}c^\dagger_{{\bf k}_3,\bar\sigma}c_{{\bf \bar q},\bar\sigma} G_W W_{{\bf \bar q},{\bf k}_4,{\bf k}_1,{\bf q}} c^\dagger_{{\bf \bar q},\bar\sigma} c_{{\bf k}_4,\bar\sigma} c^\dagger_{{\bf k}_1,\sigma} c_{{\bf q},\sigma} \\
							  &= -\sum_{1,2,3,4}\sum_\sigma c^\dagger_{{\bf k}_1\sigma} c_{{\bf k}_2\sigma} c^\dagger_{{\bf k}_3\bar\sigma} c_{{\bf k}_4\bar \sigma} \sum_{\bf q \in \text{PS}} W_{{\bf q},{\bf k}_2,{\bf k}_3,{\bf \bar q}} G_W W_{{\bf \bar q},{\bf k}_4,{\bf k}_1,{\bf q}}~,
\end{aligned}\end{equation}
\vspace{0.25cm}
\par\noindent\textbf{Scattering involving both spin-flip and spin-preserving processes}\\
%\subsubsection*{Scattering involving both spin-flip and spin-preserving processes}
These processes involve the combination of terms like \(T_{P1}\) with \(T_{F4}\), and \(T_{P2}\) with \(T_{F1}\). These again cancel each other out for the same reasons as outline above.
\vspace{0.25cm}
\par\noindent\textbf{Net renormalisation for the bath correlation term}\\
%\subsubsection*{Net renormalisation for the bath correlation term}
Since all the contributions cancel out in pairs, the bath correlation term \(W\) is {\it marginal}.

\subsection{Renormalisation of the Kondo scattering term {\it J}}
We focus on the renormalisation of the spin-flip part of the Kondo interaction. For these processes, the intermediate many-body state always involves the impurity spin being anti-correlated with the conduction electron spin, such that the propagator for that state is \(G_J = 1/\left(\omega - |\varepsilon_j|/2 + J_{\bf q}/4 + W_{\bf q}/2)\right) \).
\vspace{0.25cm}
\par\noindent\textbf{Impurity-mediated spin-flip scattering purely through Kondo-like processes}\\
%\subsubsection*{Impurity-mediated spin-flip scattering purely through Kondo-like processes}
The following processes arising from the Kondo term renormalise the spin-flip interaction:
\begin{equation}\begin{aligned}\label{kondoRenorm1}
	T^\dagger_{KT1} G_J \left(T_{KZ1} + T_{KZ1}^\dagger\right) &= \frac{1}{4}\sum_{{\bf k}_1, {\bf k}_1, {\bf q}}~J_{{\bf q}, {\bf k}_2} S_d^+ \left[-c^\dagger_{{\bf q}\downarrow} c_{{\bf k}_2 \uparrow} G_J c^\dagger_{{\bf k}_1 \downarrow} c_{{\bf q} \downarrow} + c^\dagger_{{\bf k}_2\downarrow} c_{{\bf q} \uparrow} G_J c^\dagger_{{\bf q}\uparrow} c_{{\bf k}_1 \uparrow}\right]J_{{\bf k}_1,{\bf q}} S_d^z\\
								&= -\frac{1}{8}\sum_{{\bf k}_1, {\bf k}_1, {\bf q}}~J_{{\bf q}, {\bf k}_2} S_d^+ \left[ c^\dagger_{{\bf k}_1 \downarrow}c_{{\bf k}_2 \uparrow} G_J n_{{\bf q} \downarrow} + c^\dagger_{{\bf k}_2\downarrow} c_{{\bf k}_1 \uparrow}(1 - n_{{\bf q} \uparrow}) G_J \right]J_{{\bf k}_1,{\bf q}}\\
								&= -\frac{1}{8}\sum_{{\bf k}_1, {\bf k}_1}~ S_d^+ c^\dagger_{{\bf k}_1 \downarrow}c_{{\bf k}_2 \uparrow}\sum_{{\bf q} \in \text{PS}}\left[J_{{\bf q}, {\bf k}_2} J_{{\bf k}_1,{\bf q}} + J_{{\bf \bar q}, {\bf k}_1} J_{{\bf k}_2,{\bf \bar q}} \right]G_J~.
\end{aligned}\end{equation}
In getting the final expression, we used the sigma matrix relation \(S_d^+ S_d^z = -\frac{1}{2}S_d^+\), and absorbed the projector \(1 - n_{{\bf q} \uparrow}\) into the sum over the particle sector by replacing \(q\) with its particle-hole transformed counterpart \(\bar q\). An identical contribution is obtained by switching the sequence of processes:
\begin{equation}\begin{aligned}\label{kondoRenorm2}
	 \left(T_{KZ1} + T_{KZ1}^\dagger\right) G_J T^\dagger_{KT1} &= \frac{1}{4}\sum_{{\bf k}_1, {\bf k}_1, {\bf q}}~J_{{\bf k}_1,{\bf q}} S_d^z\left[- c^\dagger_{{\bf k}_1 \downarrow} c_{{\bf q} \downarrow}G_J c^\dagger_{{\bf q}\downarrow} c_{{\bf k}_2 \uparrow} +c^\dagger_{{\bf q}\uparrow} c_{{\bf k}_1 \uparrow} G_J c^\dagger_{{\bf k}_2\downarrow} c_{{\bf q} \uparrow} \right] J_{{\bf q}, {\bf k}_2} S_d^+ \\
								&= -\frac{1}{8}\sum_{{\bf k}_1, {\bf k}_1}~ S_d^+ c^\dagger_{{\bf k}_1 \downarrow}c_{{\bf k}_2 \uparrow}\sum_{{\bf q} \in \text{PS}}\left[J_{{\bf \bar q}, {\bf k}_2} J_{{\bf k}_1,{\bf \bar q}} + J_{{\bf q}, {\bf k}_1} J_{{\bf k}_2,{\bf q}} \right]G_J~.
\end{aligned}\end{equation}
\vspace{0.25cm}
\par\noindent\textbf{Scattering processes involving interplay between the Kondo interaction and conduction bath interaction}\\
%\subsubsection*{Scattering processes involving interplay between the Kondo interaction and conduction bath interaction}
Looking at \(T_{KT1}^\dagger\) first, we have
\begin{equation}\begin{aligned}\label{t7t6}
	T_{KT1}^\dagger G_J \left(T_{F4} + T_{F4}^\dagger\right) &= \frac{1}{2}\sum_{{\bf k}_1,{\bf k}_2,{\bf q}} J_{{\bf k}_2, {\bf q}}S_d^+ \left(c^\dagger_{{\bf q}\downarrow}c_{{\bf k}_2\uparrow}  G_J W_{{\bf q}, {\bf q}, {\bf k}_1, {\bf q}} n_{q \uparrow} c^\dagger_{{\bf k}_1 \downarrow} c_{{\bf q} \downarrow} + c^\dagger_{{\bf k}_2\downarrow}c_{{\bf q}\uparrow}  G_J W_{{\bf \bar q}, {\bf \bar q}, {\bf q}, {\bf k}_1} n_{\bar q \downarrow} c^\dagger_{{\bf q} \uparrow} c_{{\bf k}_1 \uparrow}\right) ~.
\end{aligned}\end{equation}
For either of the two choices of the functional form of \(W\), it is easy to show that \(W_{{\bf q}, {\bf q}, {\bf k}_1, {\bf q}} = W_{{\bf \bar q}, {\bf \bar q}, {\bf q}, {\bf k}_1}\).
\begin{equation}\begin{aligned}\label{KT1F4}
	T_{KT1}^\dagger G_J \left(T_{F4} + T_{F4}^\dagger\right) &= \frac{1}{2}\sum_{{\bf k}_1,{\bf k}_2,{\bf q}} J_{{\bf k}_2, {\bf q}} W_{{\bf q}, {\bf q}, {\bf k}_1, {\bf q}} G_J S_d^+ \left[-c^\dagger_{{\bf k}_1 \downarrow}c_{{\bf k}_2\uparrow} n_{{\bf q}\downarrow} n_{q \uparrow} + c^\dagger_{{\bf k}_2\downarrow} c_{{\bf k}_1 \uparrow} (1 - n_{{\bf q}\uparrow}) n_{\bar q \downarrow}\right] ~.
\end{aligned}\end{equation}

Another contribution is obtained by switching the sequence of the scattering processes:
\begin{equation}\begin{aligned}\label{F4KT1}
	\left(T_{F4} + T_{F4}^\dagger\right) G_J T_{KT1}^\dagger &= \frac{1}{2}\sum_{{\bf k}_1,{\bf k}_2,{\bf q}} \left(W_{{\bf q}, {\bf q}, {\bf k}_1, {\bf q}} n_{\bar q \uparrow} c^\dagger_{{\bf k}_1 \downarrow} c_{{\bf q} \downarrow} G_J c^\dagger_{{\bf q}\downarrow}c_{{\bf k}_2\uparrow} + W_{{\bf \bar q}, {\bf \bar q}, {\bf q}, {\bf k}_1} n_{q \downarrow} c^\dagger_{{\bf q} \uparrow} c_{{\bf k}_1 \uparrow} G_J c^\dagger_{{\bf k}_2\downarrow}c_{{\bf q}\uparrow} \right) J_{{\bf k}_2, {\bf q}}S_d^+\\
															 &= \frac{1}{2}\sum_{{\bf k}_1,{\bf k}_2,{\bf q}} \left(c^\dagger_{{\bf k}_1 \downarrow} c_{{\bf k}_2\uparrow} n_{\bar q \uparrow} (1 - n_{{\bf q} \downarrow}) - c^\dagger_{{\bf k}_2\downarrow}c_{{\bf k}_1 \uparrow}n_{q \downarrow} n_{{\bf q} \uparrow}\right) W_{{\bf q}, {\bf q}, {\bf k}_1, {\bf q}} G_J J_{{\bf k}_2, {\bf q}}S_d^+
\end{aligned}\end{equation}
The two contributions (eqs.~\ref{KT1F4} and \ref{F4KT1}) arising from \(T_{KT1}\) cancel each other.

We now consider the other spin-exchange process \(T_{KT2}^\dagger\). One such contribution is
\begin{equation}\begin{aligned}\label{kondoRenorm3}
	T_{KT2}^\dagger G_J T_{F3} &= \frac{1}{2}\sum_{{\bf k}_1,{\bf k}_2,{\bf q}} J_{{\bf q}, {\bf \bar q}}S_d^+ \left(c^\dagger_{{\bf q}\downarrow}c_{{\bf \bar q}\uparrow} G_J c^\dagger_{{\bf \bar q} \uparrow} c_{{\bf k}_2 \uparrow} c^\dagger_{{\bf k}_1 \downarrow} c_{{\bf q} \downarrow} + c^\dagger_{{\bf \bar q}\downarrow}c_{{\bf q}\uparrow}  G_J  c^\dagger_{{\bf q} \uparrow} c_{{\bf k}_2 \uparrow} c^\dagger_{{\bf k}_1 \downarrow} c_{{\bf \bar q} \downarrow}\right)W_{{\bf \bar q}, {\bf k}_2, {\bf k}_1, {\bf q}}\\
							   &= -\frac{1}{2}\sum_{{\bf k}_1,{\bf k}_2,{\bf q}} S_d^+ c^\dagger_{{\bf k}_1 \downarrow} c_{{\bf k}_2 \uparrow}\left[n_{{\bf q}\downarrow}(1 - n_{{\bf \bar q}\uparrow}) + n_{{\bf \bar q}\downarrow}(1 - n_{{\bf q}\uparrow}) \right] J_{{\bf q}, {\bf \bar q}} G_J W_{{\bf \bar q}, {\bf k}_2, {\bf k}_1, {\bf q}}\\
							   &= -\frac{1}{2}\sum_{{\bf k}_1,{\bf k}_2} S_d^+ c^\dagger_{{\bf k}_1 \downarrow} c_{{\bf k}_2 \uparrow}\sum_{{\bf q} \in \text{PS}}\left(J_{{\bf q}, {\bf \bar q}} W_{{\bf \bar q}, {\bf k}_2, {\bf k}_1, {\bf q}} + J_{{\bf \bar q}, {\bf q}} W_{{\bf q}, {\bf k}_2, {\bf k}_1, {\bf \bar q}}\right)G_J~.
\end{aligned}\end{equation}
An identical contribution is obtained from the reversed term:
\begin{equation}\begin{aligned}\label{kondoRenorm4}
	T_{F3}G_J T_{KT2}^\dagger  &= \frac{1}{2}\sum_{{\bf k}_1,{\bf k}_2,{\bf q}} W_{{\bf \bar q}, {\bf k}_2, {\bf k}_1, {\bf q}}\left(c^\dagger_{{\bf \bar q} \uparrow} c_{{\bf k}_2 \uparrow} c^\dagger_{{\bf k}_1 \downarrow} c_{{\bf q} \downarrow} G_J c^\dagger_{{\bf q}\downarrow}c_{{\bf \bar q}\uparrow} +  c^\dagger_{{\bf q} \uparrow} c_{{\bf k}_2 \uparrow} c^\dagger_{{\bf k}_1 \downarrow} c_{{\bf \bar q} \downarrow} G_J c^\dagger_{{\bf \bar q}\downarrow}c_{{\bf q}\uparrow} \right)J_{{\bf q}, {\bf \bar q}}S_d^+ \\
							   &= -\frac{1}{2}\sum_{{\bf k}_1,{\bf k}_2} S_d^+ c^\dagger_{{\bf k}_1 \downarrow} c_{{\bf k}_2 \uparrow}\sum_{{\bf q} \in \text{PS}}\left(J_{{\bf q}, {\bf \bar q}} W_{{\bf \bar q}, {\bf k}_2, {\bf k}_1, {\bf q}} + J_{{\bf \bar q}, {\bf q}} W_{{\bf q}, {\bf k}_2, {\bf k}_1, {\bf \bar q}}\right)G_J~.
\end{aligned}\end{equation}
\vspace{0.25cm}
\par\noindent\textbf{Net renormalisation to the Kondo interaction}\\
%\subsubsection*{Net renormalisation to the Kondo interaction}
Combining the results from eqs.~\ref{kondoRenorm1}, \ref{kondoRenorm2}, \ref{kondoRenorm3} and \ref{kondoRenorm4}, as well as using the properties \(J_{{\bf \bar q}, {\bf k}_1} J_{{\bf k}_2,{\bf \bar q}} = J_{{\bf q}, {\bf k}_2} J_{{\bf k}_1,{\bf q}} = J_{{\bf k}_2, {\bf q}} J_{{\bf q},{\bf k}_1}\) and \(J_{{\bf q}, {\bf \bar q}} W_{{\bf \bar q}, {\bf k}_2, {\bf k}_1, {\bf q}} = J_{{\bf \bar q}, {\bf q}} W_{{\bf q}, {\bf k}_2, {\bf k}_1, {\bf \bar q}}\), the total renormalisation in the momentum-resolved Kondo coupling \(J^{(j)}\) at the \(j^\text{th}\) step amounts to
\begin{equation}\begin{aligned}
	\Delta J^{(j)}_{{\bf k}_1, {\bf k}_2} = -\sum_{{\bf q} \in \text{PS}} \frac{J^{(j)}_{{\bf k}_2,{\bf q}} J^{(j)}_{{\bf q},{\bf k}_1} + 4J^{(j)}_{{\bf q}, {\bf \bar q}} W_{{\bf \bar q}, {\bf k}_2, {\bf k}_1, {\bf q}}}{\omega - \frac{1}{2}|\varepsilon_j| + J^{(j)}_{{\bf q}}/4 + W_{{\bf q}}/2}
\end{aligned}\end{equation}

\begin{figure}
    \centering
    \includegraphics[width=0.5\linewidth]{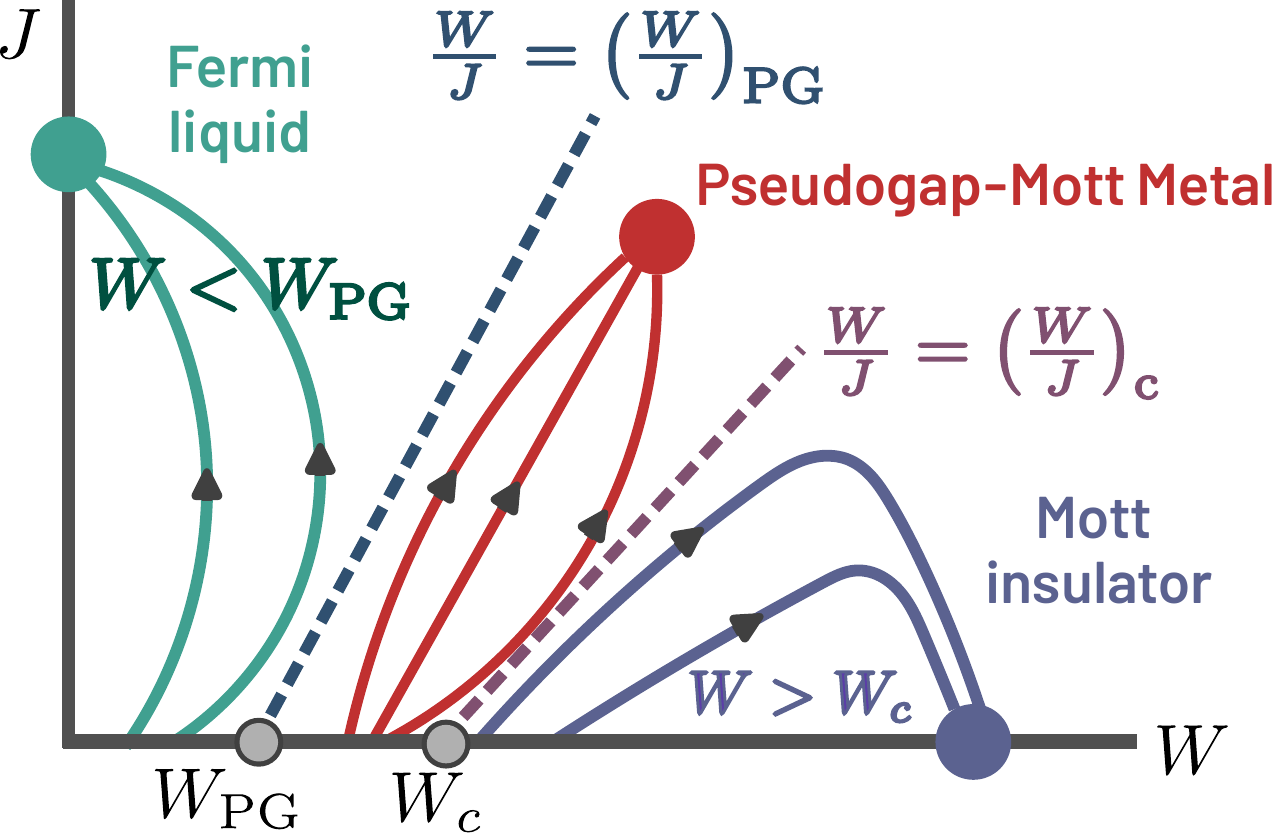}
    \caption{Renormalisation group (RG) flow diagram of the lattice model as obtained from a unitary RG analysis of the lattice-embedded impurity model. For $|W/J| < |W/J|_\text{PG}$, the flows lead to a Fermi liquid phase, while for $|W/J| < |W/J|_\text{c}$, they lead to a Mott insulator. For values in between, the system flows to a Mott metal phase that has a pseudogap in the density of states.}
    \label{rgflows}
\end{figure}

\section{Tiling formalism: Mapping from the auxiliary model to the tiled lattice model}
The eigenstates \(\ket{\Psi_{s}}\) (\(s = \left({\bf k}, n\right)\)) of the lattice Hamiltonians obtained using  the tiling procedure outlined in the main text enjoy a {\it many-body} Bloch's theorem~[53],
%\cite{stoyanova}, 
because the tiling procedure restores the translation symmetry of the Hamiltonian. This means that the {\it local} eigenstates \(\ket{\psi_n\left({\bf r}_d\right)}\) (with the impurity located at an arbitrary position \({\bf r}_d\)) of the unit cell auxiliary model Hamiltonian \(\mathcal{H}_\text{aux}({\bf r}_d)\) can be used to construct eigenstates of the lattice Hamiltonian. The index \(n(=0,1,\ldots)\) in the subscript indicates that it is the \(n^\text{th}\) eigenstate of the auxiliary model.

The state \(\ket{\psi_n\left({\bf r}_d\right)}\) does not specify the position of the zeroth site, because the unit cell Hamiltonian \(\mathcal{H}_\text{aux}({\bf r}_d)\) itself has been averaged over \(\mathcal{Z}\) zeroth sites. Accordingly, we can express the averaged eigenstate \(\ket{\psi_n\left({\bf r}_d\right)}\) as
\begin{equation}\begin{aligned}
	\ket{\psi_n\left({\bf r}_d\right)} = \frac{1}{\sqrt\mathcal{Z}}\sum_{{\bf z} \in \text{NN}({\bf r}_d)}\ket{\psi_n\left({\bf r}_d, {\bf z}\right)}~,
\end{aligned}\end{equation}
where \(\ket{\psi_n\left({\bf r}_d, {\bf z}\right)}\) is an auxiliary model eigenstate with the impurity and zeroth sites placed at \({\bf r}_d\) and \({\bf z}\). With this in mind, the following unnormalised combination of the auxiliary model eigenstates satisfies a many-particle equivalent of Bloch's theorem~[53]:
%\cite{stoyanova}:
\begin{equation}\begin{aligned}\label{eigenstateProposal1}
	\ket{\Psi_{s}} \equiv \ket{\Psi_{{\bf k}, n}} &= \frac{1}{\sqrt N}\sum_{{\bf r}_d} e^{i {\bf k}\cdot{\bf r}_d} \ket{\psi_{n}\left({\bf r}_d\right)} = \frac{1}{\sqrt{\mathcal{Z} N}}\sum_{{\bf r}_d}\sum_{{\bf z} \in \text{NN}({\bf r}_d)} e^{i {\bf k}\cdot{\bf r}_d} \ket{\psi_{n}\left({\bf r}_d, {\bf z}\right)}~,
\end{aligned}\end{equation}
where \(N\) is the total number of lattice sites and \({\bf r}_d\) is summed over all lattice spacings. The set of \(n=0\) states form the lowest band in the spectrum of the lattice, while higher values of \(n\) produce the more energetic bands. The ground state \(s = s_0\) is obtained by setting \({\bf k}\) and \(n\) to 0:
\begin{equation}\begin{aligned}\label{groundstateProposal1}
	\ket{\Psi_\text{gs}} \equiv \ket{\Psi_{s_0}} &= \frac{1}{\sqrt N}\sum_{{\bf r}_d} e^{i {\bf k}\cdot{\bf r}_d} \ket{\psi_\text{gs}\left({\bf r}_d\right)} = \frac{1}{\sqrt{\mathcal{Z} N}}\sum_{{\bf r}_d}\sum_{{\bf z} \in \text{NN}({\bf r}_d)} e^{i {\bf k}\cdot{\bf r}_d} \ket{\psi_\text{gs}\left({\bf r}_d, {\bf z}\right)}
\end{aligned}\end{equation}

The retarded time-domain lattice \(k-\)space Greens function is defined as
\begin{equation}\begin{aligned}
	\tilde G({\bf K}\sigma; t) = -i\theta(t) \braket{\Psi_\text{gs} | \left\{ c_{{\bf K}\sigma}(t), c^\dagger_{{\bf K}\sigma} \right\} | \Psi_\text{gs}}~.
\end{aligned}\end{equation}
where the bulk Hamiltonian \(H_\text{tiled}\) leads to the dynamics of the annihilation operators at time \(t\): 
\begin{equation}\begin{aligned}\label{heisenberg}
	c_{{\bf K}\sigma}(t) = e^{it H_\text{tiled} }c_{{\bf K}\sigma}e^{-i t H_\text{tiled}}~.
\end{aligned}\end{equation}
We now proceed to simplify one of the terms of the anticommutator (for simplicity of notation):
\begin{equation}\begin{aligned}\label{greensFunction1}
	&\braket{\Psi_\text{gs} | c_{{\bf K}\sigma}(t) c^\dagger_{{\bf K}\sigma} | \Psi_\text{gs}} = \frac{1}{N^2}\sum_{\vec r,\vec \Delta}e^{-i{\bf K}_0\cdot\vec\Delta}\braket{\psi_{0}(\vec r+\vec \Delta) | c_{{\bf K}\sigma}(t) c^\dagger_{{\bf K}\sigma} | \psi_{0}(\vec r)}~.
\end{aligned}\end{equation}
Using eq.~\ref{eigenstateProposal1} and the identity resolution \(1 = \sum_s\ket{\Psi_s}\bra{\Psi_s}\), eq.~\ref{greensFunction1} becomes
	\begin{equation}\begin{aligned}\label{greensfunction2}
		\braket{\Psi_\text{gs} | c_{{\bf K}\sigma}(t) c^\dagger_{{\bf K}\sigma} | \Psi_\text{gs}} = \frac{1}{N^2}\sum_s\sum_{\vec r,\vec \Delta}\sum_{\vec r^\prime,\vec \Delta^\prime}e^{-i\vec k_0\cdot\vec\Delta} e^{i\vec k\cdot\vec\Delta^\prime}\braket{\psi_{0}(\vec r+\vec \Delta) | c_{{\bf K}\sigma}(t) \ket{\psi_n(\vec r^\prime+\vec \Delta^\prime)}\bra{\psi_n(\vec r^\prime)} c^\dagger_{{\bf K}\sigma} | \psi_{0}(\vec r)}~.
\end{aligned}\end{equation}
In order to bring this expression closer to the form of an auxiliary model Greens function, we
\begin{itemize}
    \item use the relation $\ket{\psi(\vec r + \Delta)} = T^\dagger(\vec \Delta)\ket{\psi(\vec r)}$, where \(T^\dagger(\vec \Delta)\) translates all lattice sites by the vector \(\vec\Delta\),
    \item use the property $T\left(\vec a\right) c({\bf K}) T^\dagger\left(\vec a\right) = e^{-i{\bf K}\cdot\vec a}c({\bf K})$,
    \item make the substitution \(\vec \Delta^\prime \to \vec \Delta^\prime + \vec \Delta\).
\end{itemize}
This leads to the expression
\begin{equation}\begin{aligned}\label{greensfunction3}
		\braket{\Psi_\text{gs} | c_{{\bf K}\sigma}(t) c^\dagger_{{\bf K}\sigma} | \Psi_\text{gs}} = \frac{1}{N}\sum_n\sum_{\vec r,\vec r^\prime,\vec \Delta^\prime} e^{i\left( \vec k_0 + {\bf K} \right) \cdot\vec\Delta^\prime}\braket{\psi_{0}(\vec r) | c_{{\bf K}\sigma}(t) \ket{\psi_n(\vec r^\prime+\vec \Delta^\prime)}\bra{\psi_n(\vec r^\prime)} c^\dagger_{{\bf K}\sigma} | \psi_{0}(\vec r)}~,
\end{aligned}\end{equation}
where the sum over \(s=(\vec k, n)\) has been reduced to a sum over the auxiliary model eigenstate index \(n\) because of the Kronecker delta \(\delta\left(\vec k_0 + {\bf K} - \vec k\right)\). This can be further simplified by splitting the sum over \(\vec\Delta^\prime\) into positive and negative parts and then making the transformation \(\vec r^\prime \to \vec r^\prime + \vec \Delta^\prime\):
\begin{equation}\begin{aligned}
	\sum_{\vec r^\prime,\vec \Delta^\prime} e^{i\left( \vec k_0 + {\bf K} \right) \cdot\vec\Delta^\prime}\ket{\psi_n(\vec r^\prime+\vec \Delta^\prime)}\bra{\psi_n(\vec r^\prime)} &= \frac{1}{2}\sum_{\vec r^\prime,\vec \Delta^\prime}\left[ e^{i\left( \vec k_0 + {\bf K} \right) \cdot\vec\Delta^\prime}\ket{\psi_n(\vec r^\prime+\vec \Delta^\prime)}\bra{\psi_n(\vec r^\prime)} + e^{-i\left( \vec k_0 + {\bf K} \right) \cdot\vec\Delta^\prime}\ket{\psi_n(\vec r^\prime-\vec \Delta^\prime)}\bra{\psi_n(\vec r^\prime)}\right] \\
																																												   &= \frac{1}{2}\sum_{\vec r^\prime,\vec \Delta^\prime}\left[ e^{i\left( \vec k_0 + {\bf K} \right) \cdot\vec\Delta^\prime}\ket{\psi_n(\vec r^\prime+\vec \Delta^\prime)}\bra{\psi_n(\vec r^\prime)} + e^{-i\left( \vec k_0 + {\bf K} \right) \cdot\vec\Delta^\prime}\ket{\psi_n(\vec r^\prime)}\bra{\psi_n(\vec r^\prime+\vec \Delta^\prime)}\right] \\
\end{aligned}\end{equation}
For each pair of \(\vec r^\prime\) and \(\vec\Delta^\prime\), the term within the box brackets has the form of a two-level Hamiltonian between the states \(\ket{\psi_n(\vec r^\prime)}\) and \(\ket{\psi_n(\vec r^\prime+\vec \Delta^\prime)}\), with a tunnelling amplitude \(e^{i\left( \vec k_0 + {\bf K} \right) \cdot\vec\Delta^\prime}\). The term can therefore be written in the eigenbasis of this Hamiltonian:
\begin{equation}\begin{aligned}
	\sum_{\vec r^\prime,\vec \Delta^\prime} e^{i\left( \vec k_0 + {\bf K} \right) \cdot\vec\Delta^\prime}\ket{\psi_n(\vec r^\prime+\vec \Delta^\prime)}\bra{\psi_n(\vec r^\prime)} &=\frac{1}{2}\sum_{\vec r^\prime,\vec \Delta^\prime}\left[ \ket{\chi_n^+(\vec r^\prime,\vec\Delta^\prime)}\bra{\chi_n^+(\vec r^\prime,\vec\Delta^\prime)} - \ket{\chi_n^-(\vec r^\prime,\vec\Delta^\prime)}\bra{\chi_n^-(\vec r^\prime,\vec\Delta^\prime)}\right],
\end{aligned}\end{equation}
where \(\ket{\chi_n^\pm(\vec r^\prime,\vec\Delta^\prime)} = \frac{1}{\sqrt 2}\left[\ket{\psi_n(\vec r^\prime)} \pm e^{i\left( \vec k_0 + {\bf K} \right) \cdot\vec\Delta^\prime}\ket{\psi_n(\vec r^\prime + \vec\Delta^\prime)}\right] \) are the eigenvectors of the tunnelling Hamiltonian with eigenvalues \(\pm 1\) respectively. With this basis transformation, we can rewrite eq.~\ref{greensfunction3} as
\begin{equation}\begin{aligned}
		\braket{\Psi_\text{gs} | c_{{\bf K}\sigma}(t) c^\dagger_{{\bf K}\sigma} | \Psi_\text{gs}} = \frac{1}{2N}\sum_n\sum_{\vec r,\vec r^\prime,\vec \Delta^\prime} \braket{\psi_{0}(\vec r) | c_{{\bf K}\sigma}(t) \left[ \ket{\chi_n^+(\vec r^\prime,\vec\Delta^\prime)}\bra{\chi_n^+(\vec r^\prime,\vec\Delta^\prime)} - \ket{\chi_n^-(\vec r^\prime,\vec\Delta^\prime)}\bra{\chi_n^-(\vec r^\prime,\vec\Delta^\prime)}\right] c^\dagger_{{\bf K}\sigma} | \psi_{0}(\vec r)}~.
\end{aligned}\end{equation}

In the present work, we consider only the $\vec r^\prime = \vec r,~\vec\Delta^\prime=0$ component. These terms represent those contributions to the total Greens function that arise from excitations that start and end at a specific auxiliary model (at \(\vec r\)), and also evolve dynamically within the same auxiliary model. These terms are therefore exactly equal to the auxiliary model Greens function at position \(\vec r\), and are the most dominant contribution due to the localised nature of the impurity model.

Restricting ourselves to just the single auxiliary model contributions gives
\begin{equation}\begin{aligned}
		\braket{\Psi_\text{gs} | c_{{\bf K}\sigma}(t) c^\dagger_{{\bf K}\sigma} | \Psi_\text{gs}} = \frac{1}{N}\sum_n\sum_{\vec r} \braket{\psi_{0}(\vec r) | c_{{\bf K}\sigma}(t) \ket{\psi_n(\vec r)}\bra{\psi_n(\vec r)} c^\dagger_{{\bf K}\sigma} | \psi_{0}(\vec r)} ~.
\end{aligned}\end{equation}
We first consider more carefully the transition operator \(\mathcal{T}_{{\bf K}\sigma} = c_{{\bf K}\sigma}\) for the 1-particle excitation giving rise to the above Greens function. Within our auxiliary model approach, gapless excitations within the lattice model are represented by gapless excitations of the impurity site, specifically those that screen the impurity site and form the local Fermi liquid. As a result, the uncoordinated \(\mathcal{T}-\)matrix for the lattice model must be replaced by a combined \(\mathcal{T}-\)matrix within the impurity model that captures those gapless excitations that occur in connection with the impurity, and projects out the uncorrelated excitations that take place even when the impurity site is decoupled from the bath.

In order to construct this auxiliary model \(\mathcal{T}-\)matrix, we note that the impurity site can have both spin and charge excitations. Considering both excitations, the modified \(\mathcal{T}-\)matrix that constructs \(k-\)space excitations in correlation with the impurity site are
\begin{equation}\begin{aligned}\label{tmatrix}
	\mathcal{T}_{{\bf K}\sigma} = c_{{\bf K}\sigma}\left(\sum_{\sigma^\prime}c^\dagger_{d\sigma} + \text{h.c.}\right) + c_{{\bf K}\sigma}\left(S_d^+ + \text{h.c.}\right)~,
\end{aligned}\end{equation}
leading to the updated expression for the complete Greens function:
\begin{equation}\begin{aligned}
	\tilde G({\bf K}\sigma; t) = -i\theta(t)\frac{1}{N}\sum_n\sum_{\vec r} \braket{\psi_{0}(\vec r) | \left[\mathcal{T}_{{\bf K}\sigma}(t) \ket{\psi_n(\vec r)}\bra{\psi_n(\vec r)} \mathcal{T}_{{\bf K}\sigma}^\dagger + \mathcal{T}^\dagger_{{\bf K}\sigma} \ket{\psi_n(\vec r)}\bra{\psi_n(\vec r)} \mathcal{T}_{{\bf K}\sigma}(t)\right] | \psi_{0}(\vec r)} ~.
\end{aligned}\end{equation}

\section{Theory for the nodal non-Fermi liquid: effective Hatsugai-Kohmoto model}
As the bath interaction \(W\) is tuned through the L-PG phase, the nodal region is the last to decouple from the impurity. This allows us to write down a simpler Kondo model near the transition, where only the nodal region is hybridising with the impurity spin through Kondo interactions. This is done by retaining only those scattering processes \({\bf k}_1 \to {\bf k}_2\) that originate from and end at \(k-\)points within a small neighborhood of width \(|{\bf q}|\) around the four nodal points: \({\bf k}_1, {\bf k}_2 \in {\bf N} + {\bf q}, |{\bf q}| \ll \pi\), where \({\bf N}\) can be any one of the four nodal points \({\bf N}_{1} = \left(\pi/2, \pi/2\right), {\bf N}_{2} = \left(-\pi/2, \pi/2\right)\) and \({\bf N}_1 + {\bf Q}_1\) and \({\bf N}_{2} + {\bf Q}_2\), where \({\bf Q}_1 = \left(-\pi,-\pi \right)\) and \({\bf Q}_2 = \left( \pi, -\pi \right) \) are the two nesting vectors. We assume that the window of \({\bf q}\) is small enough so that the fixed point Kondo coupling values \(J^*({\bf q}_1, {\bf q}_2)\) for the scattering processes involving \({\bf q}_1\) and \({\bf q}_2\) can be replaced by an average value \(J^*\).

With these considerations, the simplified low-energy model near the transition describing the Kondo scattering processes can be written as
\begin{equation}\begin{aligned}
	\tilde H_\text{imp-cbath} = J^*\frac{1}{2}\sum_{l=1,2}\sum_{{\bf q}_1, {\bf q}_2}\sum_{\alpha,\beta}{\bf S}_d\cdot{\boldsymbol \sigma}_{\alpha\beta}\left(c^\dagger_{{\bf N}_l + {\bf q}_1,\alpha}c_{{\bf N}_l + {\bf q}_2,\beta} + c^\dagger_{{\bf N}_l + {\bf Q}_l + {\bf q}_1,\alpha}c_{{\bf N}_l + {\bf Q}_l + {\bf q}_2,\beta}  - c^\dagger_{{\bf N}_l + {\bf Q}_l + {\bf q}_1,\alpha}c_{{\bf N}_l + {\bf q}_2,\beta} \right.\\
    \left.- c^\dagger_{{\bf N}_l + {\bf q}_1,\alpha}c_{{\bf N}_l + {\bf Q}_l + {\bf q}_2,\beta}\right)~.
\end{aligned}\end{equation}
The label \(l\) can take values 1 or 2, allowing us to consider both the decoupled channels in the PG (associated with \({\bf N}_1\) and \({\bf N}_2\)). It also labels the nesting vectors \({\bf Q}_l\) associated with the two sets. \(\alpha\) and \(\beta\) indicate spin indices, and \({\bf q}_1\) and \({\bf q}_2\) represent incoming and outgoing momenta in the scattering processes.

Because of the decoupling of the channel \(l=1\) and \(l=2\), we consider only the \(l=1\) channel for the rest of the calculations in this section. In order to simplify the Hamiltonian, we define new fermionic operators
\begin{equation}\begin{aligned}
	\psi_{{\bf q}, \sigma} = \frac{1}{\sqrt 2}\left(c_{{\bf N}_1 + {\bf q},\sigma} + c_{{\bf N}_1 + {\bf Q}_1 - {\bf q}, \sigma}\right)~, ~\phi_{{\bf q}, \sigma} = \frac{1}{\sqrt 2}\left(c_{{\bf N}_1 + {\bf q},\sigma} - c_{{\bf N}_1 + {\bf Q}_1 - {\bf q}, \sigma}\right)~,
\end{aligned}\end{equation}
The operator satisfy fermionic anticommutation relations. For convenience, we define new number operators for the sum and relative degrees of freedom:
\begin{equation}\begin{aligned}
	s_{{\bf q},\sigma} = {\psi}^\dagger_{{\bf q}} {\psi}_{{\bf q}},~ r_{{\bf q},\sigma} = {\phi}^\dagger_{{\bf q}} {\phi}_{{\bf q}}~.
\end{aligned}\end{equation}
We then have the following useful relation between the corresponding number operators \(n = c^\dagger c\):
\begin{equation}\begin{aligned}\label{numberOperatorRelation}
	n_{{\bf N}_1 + {\bf q},\sigma} + n_{{\bf N}_1 + {\bf Q}_1 - {\bf q}, \sigma} = s_{{\bf q}, \sigma} + r_{{\bf q},\sigma}~.
\end{aligned}\end{equation}

In terms of these new degrees of freedom, the auxiliary model Hamiltonian takes the form
\begin{equation}\begin{aligned}
	\tilde H = -\frac{1}{2}W\sum_{{\bf q},\sigma}r_{{\bf q},\sigma} + W\sum_{{\bf q}_1,{\bf q}_2}r_{{\bf q}_1,\uparrow} r_{{\bf q}_2,\downarrow} + \sum_{{\bf q}_1, {\bf q}_2,\alpha,\beta}J^*{\bf S}_d\cdot{\boldsymbol\sigma}_{\alpha\beta} {\phi}^\dagger_{{\bf q}_1\alpha} \phi_{{\bf q}_2\beta} + \sum_{{\bf q},\sigma}\varepsilon_{{\bf N}_1 + {\bf q}}\left({\psi}^\dagger_{{\bf q},\sigma} {\phi}_{{\bf q},\sigma} + {\phi}^\dagger_{{\bf q},\sigma} \psi_{{\bf q},\sigma}\right),
\end{aligned}\end{equation}
where we have considered a simplified form of the bath interaction (for the channel \(l=1\)), taking into account the density-density correlations in \(k-\)space. 

For bath interaction strength close to the critical value (\(W \lesssim W_c\)), the fixed point coupling value \(J^*\) is much smaller than \(W\). In order to obtain the gapless excitations of the system arising from the presence of the impurity site, we integrate out the impurity dynamics via a Schrieffer-Wolff transformation. The perturbation term \(\mathcal{V}\) then consists of Hamiltonian terms that modify the impurity configuration,
\begin{equation}\begin{aligned}
	\mathcal{V} = \sum_{{\bf q},\sigma}\varepsilon_{{\bf N}_1 + {\bf q}}\left({\psi}^\dagger_{{\bf q},\sigma} {\phi}_{{\bf q},\sigma} + {\phi}^\dagger_{{\bf q},\sigma} \psi_{{\bf q},\sigma}\right) + \sum_{{\bf q}_1, {\bf q}_2}J^*{S}_d^+{\phi}^\dagger_{{\bf q}_1 \downarrow} \phi_{{\bf q}_2 \uparrow} + \text{h.c.}~,
\end{aligned}\end{equation}
while the ``non-interacting" Hamiltonian is
\begin{equation}\begin{aligned}
	H_D &= -\frac{1}{2}W\sum_{{\bf q},\sigma}r_{{\bf q},\sigma} + W\sum_{{\bf q}_1,{\bf q}_2}r_{{\bf q}_1,\uparrow} r_{{\bf q}_2,\downarrow}~.
\end{aligned}\end{equation}
For the present Hamiltonian, the low-energy state is the one that minimises the bath interaction term \(\tilde H_\text{cbath-int}\). High-energy states are obtained by applying, on the state \(\ket{L}\), the excitation operator \(\phi^\dagger_{{\bf q}_1,\sigma}\phi_{{\bf q}_2,\bar\sigma}\) or its hermitian conjugate.

The complete second-order renormalised Hamiltonian is
\begin{equation}\begin{aligned}
    \Delta \tilde H =& \sum_{{\bf q}, \sigma}\epsilon_{\bf q} r_{{\bf q},\sigma} + \mathcal{U}\sum_{{\bf q}, \sigma}r_{{\bf q} \sigma}r_{{\bf q} \bar\sigma} + \mathcal{U}\sum_{{\bf q}_1 \neq {\bf q}_2, \sigma}\left[r_{{\bf q}_1 \sigma}r_{{\bf q}_2 \bar\sigma} + \phi^\dagger_{{\bf q}_1,\bar\sigma}\phi^\dagger_{{\bf q}_1,\sigma}\phi_{{\bf q}_2, \sigma}\phi_{{\bf q}_2, \bar\sigma}\right]~.
\end{aligned}\end{equation}
where $\epsilon_{\bf q} = \text{sign}\left(\varepsilon_{{\bf N}_1 + {\bf q}}\right)\frac{\varepsilon_{{\bf N}_1 + {\bf q}}^2}{-W}$ and $\mathcal{U} = \frac{{J^*}^2}{4W}$.

The ${\bf q}_1={\bf q}_2$ component of the Hamiltonian shows the emergence of the exactly solvable Hatsugai-Kohmoto model~[42,43]
%\cite{Baskaran1991,Hatsugai1992} 
at the critical point. The correlation term $\mathcal{U}\sum_{{\bf q}, \sigma}r_{{\bf q} \sigma}r_{{\bf q} \bar\sigma}$ leads to a transfer of spectral weight across the Fermi surface and separates the available states into three classes:
\begin{equation}\begin{aligned}
\braket{n_{\bf q}} = \begin{cases}
    2, \quad \epsilon_{\bf q} < -|\mathcal{U}/2|~,\\
    1, \quad |\mathcal{U}/2| > \epsilon_{\bf q} > -|\mathcal{U}/2|~,\\
    0, \quad |\mathcal{U}/2| > \epsilon_{\bf q}~.
\end{cases}
\end{aligned}\end{equation}
Different from a Fermi liquid is the emergence of the highly degenerate single-occupied region in the middle, and this gives rise to non-Fermi liquid excitations.

We now discuss the effects of the ${\bf q}_1 \neq {\bf q}_2$ component. The first term partially lifts the degeneracy of the central singly-occupied region and allows only zero magnetisation configurations. The second term creates gapped excitations involving the regions of zero and double occupancy; these represent subdominant pairing fluctuations of the nodal non-Fermi liquid.

\section{Luttinger’s theorem in the presence of Luttinger surfaces.}
A natural question is whether Luttinger’s theorem continues to hold in our model in the presence of Luttinger surfaces. It has been shown that particle-hole symmetric systems satisfy a generalized version of Luttinger’s theorem~[34],
%\cite{seki2017topological}, 
wherein the Luttinger volume $V_L$ is given by the difference between the number of poles and zeros of the single-particle Green’s function enclosed by the FS. This statement remains valid even when the self-energy $\Sigma(\mathbf{k}, \omega)$ diverges~[32].
%\cite{Phillips2013}. 
In our case, particle-hole symmetry is preserved and the system remains at half-filling, ensuring the total number of occupied states remains constant. Prior to the Lifshitz transition, gapless excitations on the FS contribute one pole per momentum state to the Luttinger count. Inside the PG phase, these gapless $k$-states persist, but the emergence of gapped regions redistributes spectral weight: doubly occupied states on Luttinger surfaces become energetically favorable due to the attractive $W$-interaction. These doubly occupied states are degenerate with the empty states by particle-hole symmetry, and thus, on average, one state remains occupied. The net result is that the number of occupied states - hence the Luttinger volume - remains unchanged. This argument extends to the MI, wherein the FS is entirely replaced by a Luttinger surface with zero quasiparticle weight.

\section{Additional Results on the Fermi Liquid \& Pseudogap-Mott metal Phases}
\begin{figure}[!h]
	\centering
	\includegraphics[width=0.24\textwidth]{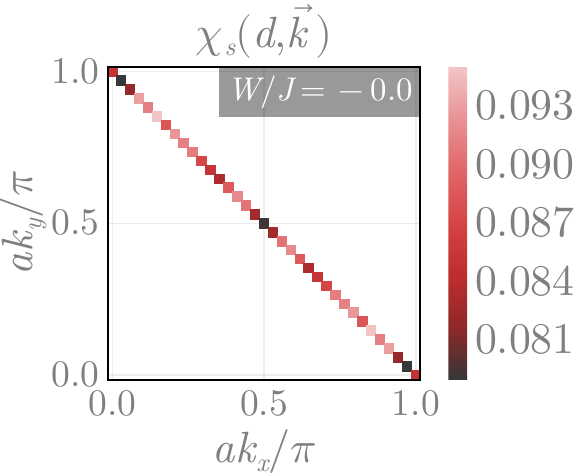}
	\includegraphics[width=0.24\textwidth]{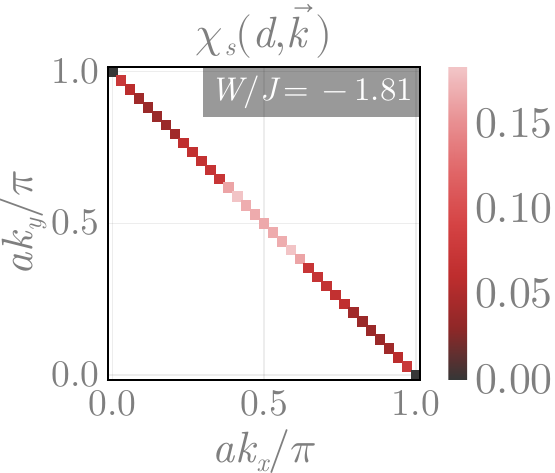}
	\includegraphics[width=0.24\textwidth]{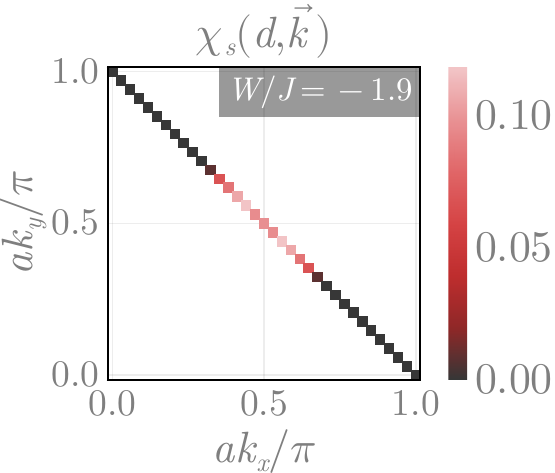}
	\includegraphics[width=0.24\textwidth]{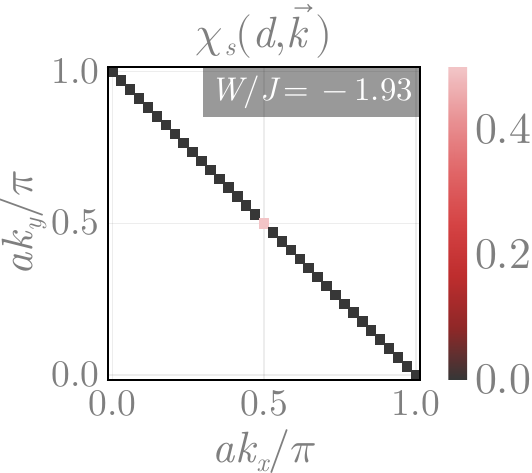}
	\caption{\(k-\)space distribution of spin-spin correlation \(\chi_s(d,\vec k)\) between the impurity spin and momentum states in the conduction bath, calculated for a $69\times 69$ $k$-space grid. The last three figures show how \(k-\)points starting from the antinode progressively exit the Kondo cloud, the node being the last \(k-\)point to decouple from the impurity.}
	\label{spinCorr}
\end{figure}

\begin{figure}[!h]
	\centering
	\includegraphics[width=0.24\textwidth]{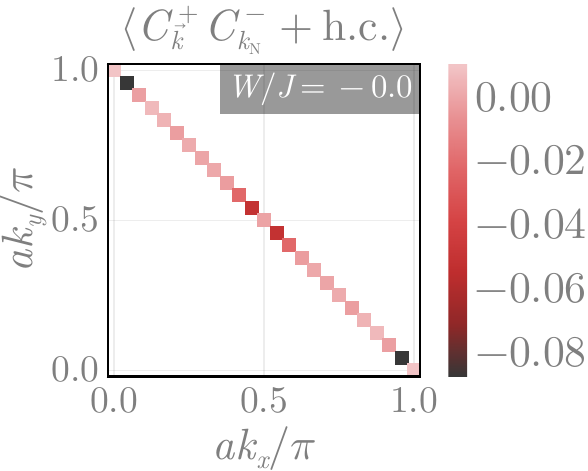}
	\includegraphics[width=0.24\textwidth]{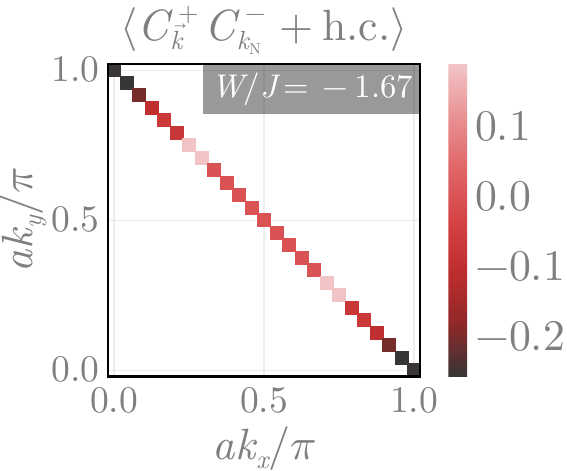}
	\includegraphics[width=0.24\textwidth]{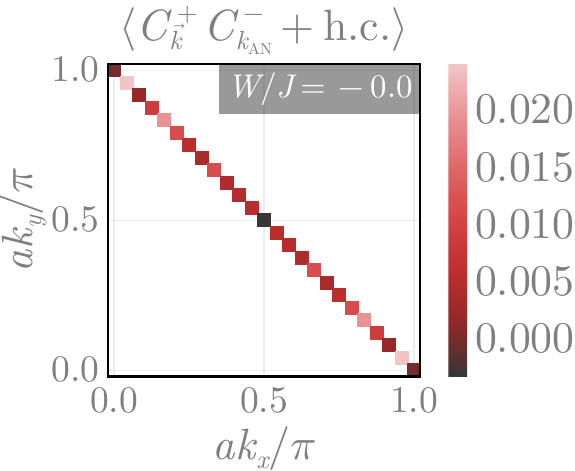}
	\includegraphics[width=0.24\textwidth]{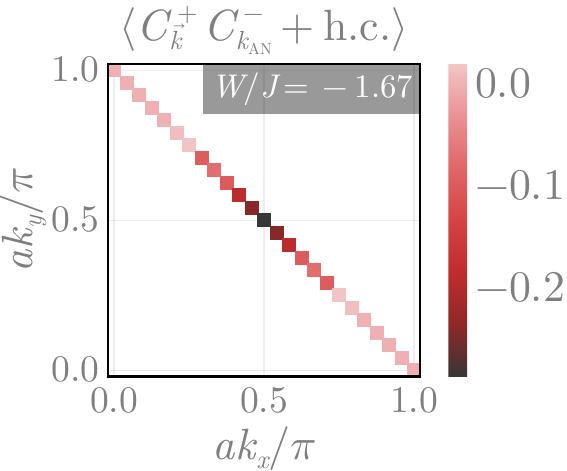}
	\caption{Charge correlation \(\chi_c\) starting from antinode (left) and node (right), at the beginning of the pseudogap phase, calculated for a $49\times 49$ $k$-space grid. Strong node-antinode correlations are clearly visible.}
	\label{cfnode}
\end{figure}

\begin{figure}
    \centering
    \includegraphics[width=0.4\linewidth]{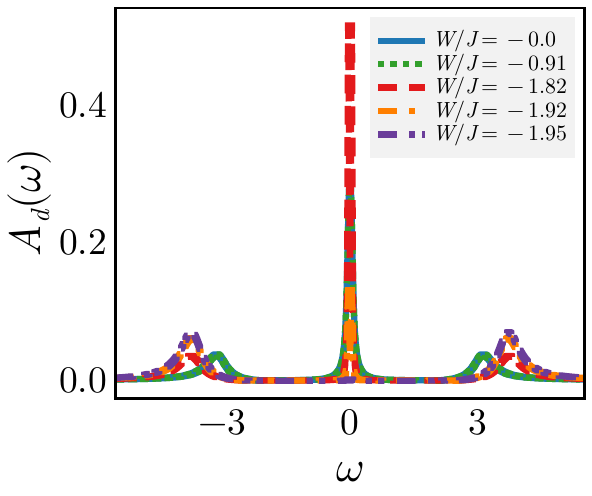}
    \caption{Impurity Spectral function in the FL and PG phases, calculated for a $77\times 77$ $k$-space grid. The evolution of the central peak from Kondo resonance to pseudogap is accompanied by dynamical spectral weight transfer to the Hubbard side bands at finite frequencies $\omega\simeq \pm 3$ (in units of the bandwidth).}
    \label{specfunc}
\end{figure}

\begin{figure}[!h]
	\centering
	\includegraphics[width=0.24\textwidth]{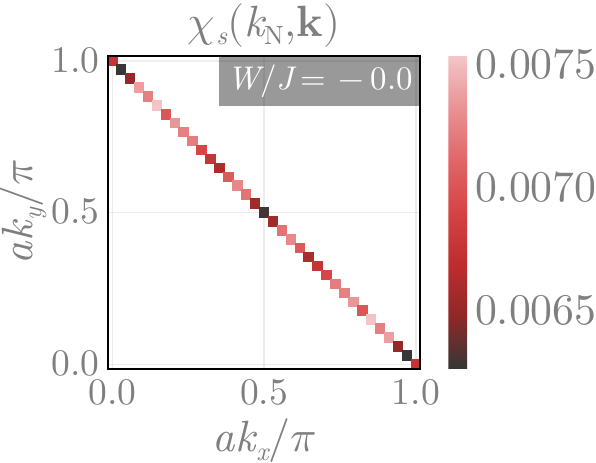}
	\includegraphics[width=0.24\textwidth]{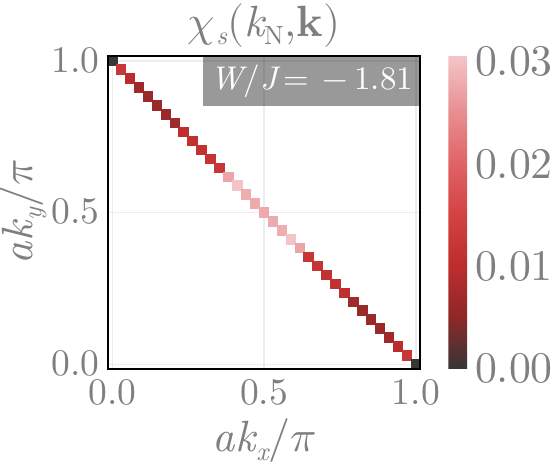}
	\includegraphics[width=0.24\textwidth]{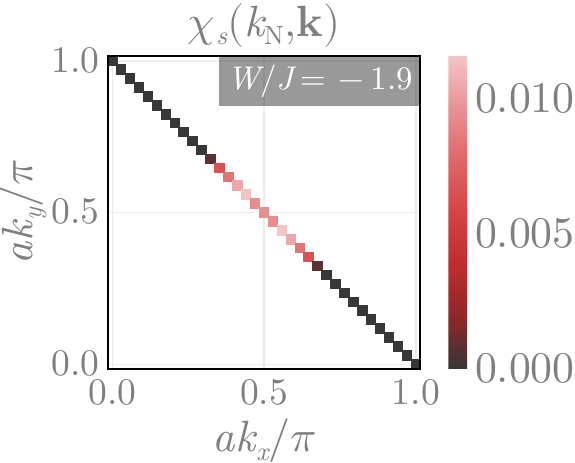}
	\includegraphics[width=0.24\textwidth]{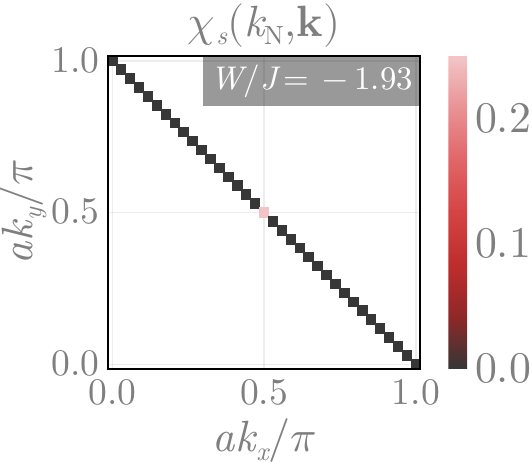}
	\caption{Spin-spin correlations \(\chi_s\) for the {\it tiled model}, between the nodal point \(k_N = \left( -\pi/2, -\pi/2\right)\) and an arbitrary \(k-\)point on the Fermi surface, calculated for a $69\times 69$ $k$-space grid. In the absence of bath interaction (first panel), the correlations are somewhat uniformly distributed along the Fermi surface, and quite small. This describes the Fermi liquid phase of the lattice model. As we enter the pseudogap (second panel and beyond), the spin-correlations near the antinode vanish, indicating that they have been removed from the metallic excitations, while the correlations near the nodal point become enhanced because of the increasingly correlated nature of the metal.}
	\label{tiledSF}
\end{figure}

\begin{figure}[!h]
	\includegraphics[width=0.24\textwidth]{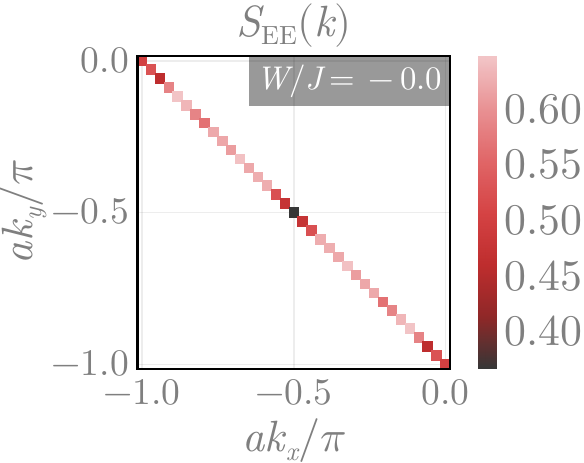}
	\includegraphics[width=0.24\textwidth]{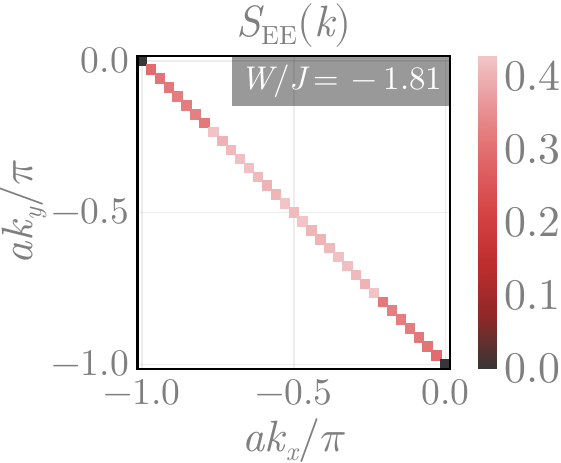}
	\includegraphics[width=0.24\textwidth]{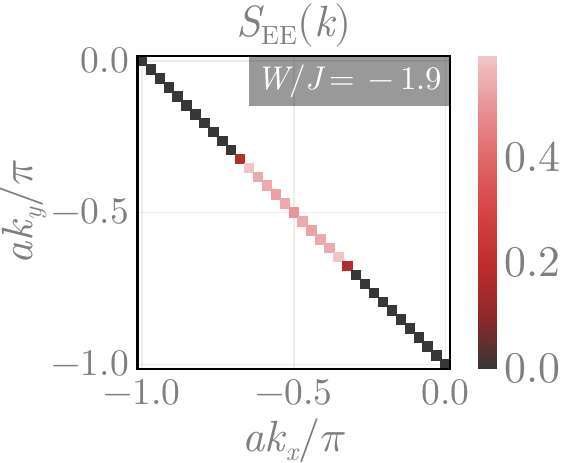}
	\includegraphics[width=0.24\textwidth]{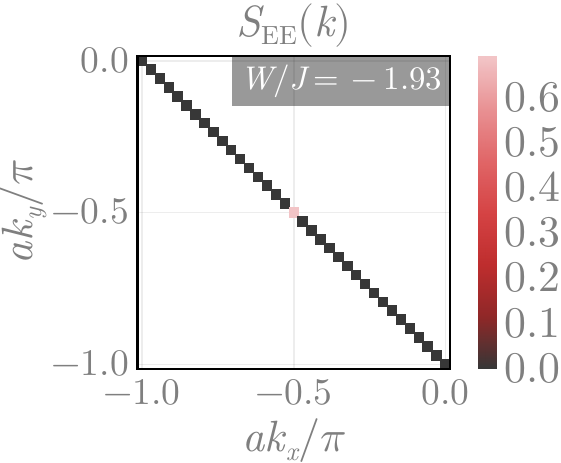}

	\includegraphics[width=0.24\textwidth]{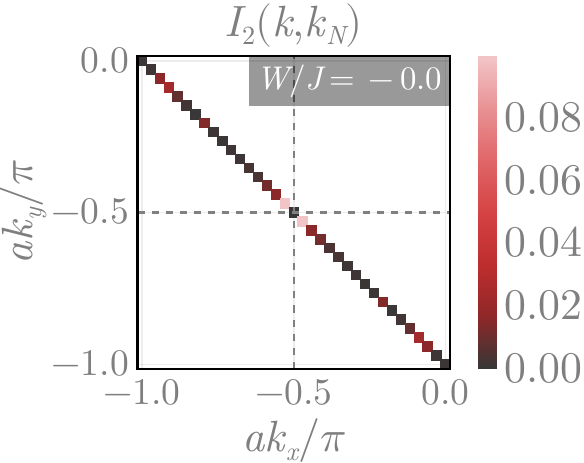}
	\includegraphics[width=0.24\textwidth]{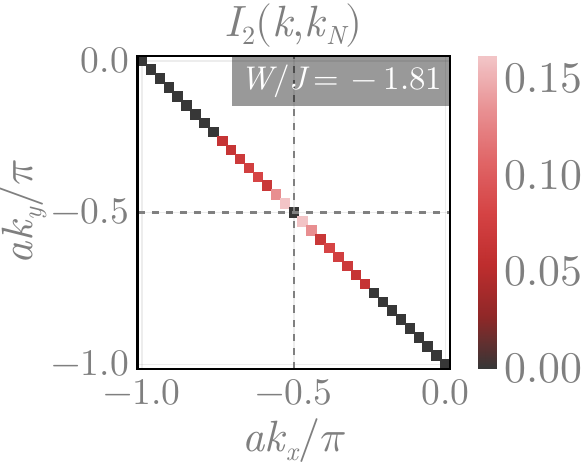}
	\includegraphics[width=0.24\textwidth]{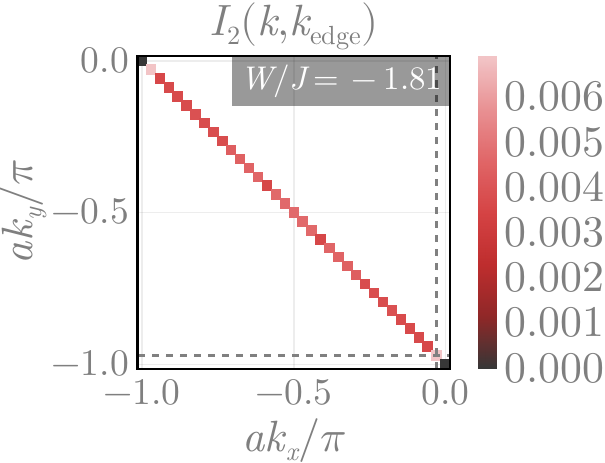}
	\includegraphics[width=0.24\textwidth]{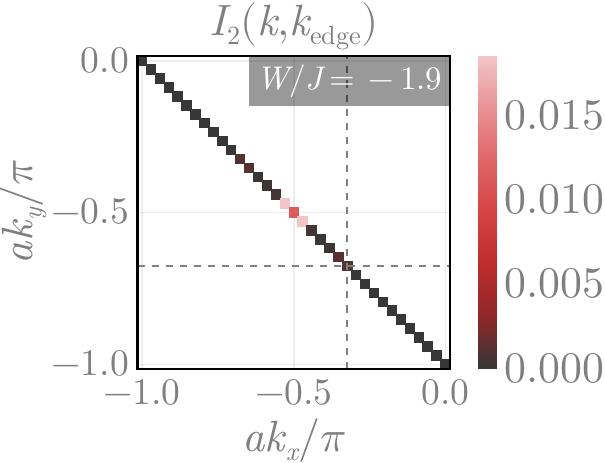}
	\caption{Top panel: Entanglement entropy \(S_\text{EE}({\bf k})\) on the Fermi surface of the tiled model, calculated for a $69\times 69$ $k$-space grid. Similar to the correlations, the entanglement in the \(W=0\) model remains uniformly spread out over the Fermi surface. As the Fermi surface progressively shrinks, the entanglement gets concentrated to the nodal points. Bottom panel: Mutual information \(I_2\) between (i) an arbitrary \(k-\)state and the nodal point (first and second plots), and (ii) an arbitrary \(k-\)state and the edge of the disconnected Fermi surface in the pseudogap (third and fourth plots). The ends of the partial Fermi surface appear to be weakly entangled to the rest of the Fermi surface, while the nodal points remain strongly entangled all the way through to the transition.}
	\label{tiledEntanglement}
\end{figure}

We present additional figures of the spin (Fig.\ref{spinCorr}) and charge (Fig.\ref{cfnode}) correlations in the impurity model, the impurity spectral function (Fig.\ref{specfunc}), the spin correlations (Fig.\ref{tiledSF}), entanglement entropy (Fig.\ref{tiledEntanglement} (upper panel)) and mutual function (Fig.\ref{tiledEntanglement} (lower panel)) of the tiled model in the pseudogap phase.

\section{Reconstructing Fermi liquid theory from local Fermi liquid excitations}
For $|W| < |W_\text{PG}|$, the impurity spin gets screened by the conduction bath at low temperatures, and the low-energy physics of the impurity model is described by local Fermi liquid excitations at the conduction bath sites nearest neighbour to the impurity~\cite{nozieres1974fermi}:
\begin{equation}\begin{aligned}
    H_\text{LFL} = \epsilon \sum_{Z,\sigma} n_{Z\sigma} + U \sum_{Z}n_{Z \uparrow}n_{Z \downarrow},\quad U > 0~,
\end{aligned}\end{equation}
where $Z$ sums over the nearest-neighbour sites. Given the purely local nature of the effective Hamiltonian and the repulsive interaction $U$, excitations are generated through the local operators $c^\dagger_{Z\sigma}$. In order to obtain the excitations of the tiled model, we apply Bloch's theorem to the excitations of the auxiliary model. To find the form of the excitation at total crystal momentum ${\bf K}$, we have
\begin{equation}\begin{aligned}
    c^\dagger_{Z\sigma} \to \sum_{{\bf r}} e^{-i {\bf K}} c^\dagger_{{\bf r}\sigma} = c^\dagger_{{\bf K},\sigma}~,
\end{aligned}\end{equation}
where the specific lattice site ${\bf Z}$ has been replaced with the translated index ${\bf r}$. The tiled excitations are therefore one-particle in nature as well, with momentum as a quantum number. This shows that local Fermi liquid excitations in the auxiliary model very naturally lead to Fermi liquid excitations in the bulk lattice model.

\section{Heisenberg Model as a low-energy description of the tiled Mott insulator}\label{heisenbergDerivation}
In the insulating phase, the ground state of each auxiliary model hosts a decoupled local moment. Upon applying the tiling procedure, the lattice model ground state becomes that of the Hubbard model in the atomic limit. In order to lift the extensive degeneracy of the state, we will now take into consideration inter-auxiliary model virtual scattering processes that were subdominant in the metallic phase and were hence ignored. These one-particle scattering processes lead to the emergence of a nearest-neighbour superexchange interaction. The calculation is a straightforward application of second order perturbation theory (Schrieffer-Wolff transformation). In order to allow virtual fluctuations that can lift the large ground state degeneracy and lower the energy, we consider (perturbatively) the effects of an irrelevant single-particle hybridisation that connects the nearest-neighbour sites.

For simplicity, we consider two impurity sites labelled \(1\) and \(2\) associated with two nearest-neighbour auxiliary models. The ground state subspace is four-fold degenerate:
\begin{equation}\begin{aligned}
	\ket{\Psi_L} = \left\{\ket{\sigma_1,\sigma_2}\right\}~,\quad \sigma_i = \pm 1~,
\end{aligned}\end{equation}
where \(\sigma_i\) is the spin state of site \(i\). This ground state is derived from the following "zeroth order" Hamiltonian that emerges in the local moment phase of the auxiliary models when all scattering processes between the impurity and conduction bath are RG-irrelevant:
\begin{equation}\begin{aligned}
	H_0 = -\frac{U}{2}\sum_{i=1,2}\left(n_{i \uparrow} - n_{i \downarrow}\right)^2~;
\end{aligned}\end{equation}
the local correlation on the impurity site becomes the largest scale in the problem in this phase and pushes the \(\ket{n_i=2}\) and \(\ket{n_i=0}\) states to high energies. This then defines the high-energy subspace for our calculation:
\begin{equation}\begin{aligned}
	\ket{\Psi_H} = \ket{C_1,C_2}~,
\end{aligned}\end{equation}
where \(C_i\) can take values 0 or 2, indicating that the state \(i\) is either empty or full, respectively. Both the double and hole states exist at a charge gap of the order of \(U/2\) above the low-energy singly-occupied subspace defined by the states \(\ket{\Psi_L}\).

In order to allow virtual fluctuations that can lift the large ground state degeneracy and lower the energy, we consider (perturbatively) the effects of an irrelevant single-particle hybridisation that connects the nearest-neighbour sites. This perturbation Hamiltonian is therefore of the form
\begin{equation}\begin{aligned}
	H_t = \sum_\omega V(\omega) \mathcal{P}(\omega)~\sum_\sigma\left(c^\dagger_{1\sigma}c_{2\sigma} + \text{h.c.}\right) ~,
\end{aligned}\end{equation}
where \(V(\omega)\) only acts on states at the energy scale \(\omega\); the renormalisation of \(V\) is encoded in the fact that \(V(\omega)\) is largest for the excited states and vanishes at low-energies: \(V(\omega \to 0) = 0\).

In order to obtain a low-energy effective Hamiltonian for the impurity sites arising from this hybridisation, we integrate out \(H_t\) via a Schrieffer-Wolff transformation. This leads to the following second-order Hamiltonian:
\begin{equation}\begin{aligned}
	H_\text{eff} = \mathcal{P}_L H_t G\mathcal H_t \mathcal{P}_L~.
\end{aligned}\end{equation}
The operator \(\mathcal{P}_L\) projects onto the low-energy subspace \(\ket{\Psi_L}\) - this ensures that we remain in the low-energy subspace at the beginning and at the end of the total process. The Greens function \(G = (E_L - H_0)^{-1}\) incorporates the excitation energy to go from the low-energy subspace \(\ket{\Psi}_L\) (of energy \(E_L\)) to the excited subspace \(\ket{\Psi}_H\) of energy \(E_L + U/2\). Substituting the form of the perturbation Hamiltonian and the excitation energy into the above expression gives
\begin{equation}\begin{aligned}\label{effHam1}
	H_\text{eff} = \frac{V_H^2}{-U/2}\sum_{\sigma,\sigma^\prime} \left[c^\dagger_{1\sigma}c_{2\sigma}c^\dagger_{2\sigma^\prime}c_{1\sigma^\prime} + c^\dagger_{2\sigma}c_{1\sigma}c^\dagger_{1\sigma^\prime}c_{2\sigma^\prime}\right] ~.
\end{aligned}\end{equation}
where \(V_H \equiv V(\omega \to U/2)\) is the impurity-bath hybridisation at energy scales of the order of the Mott gap, in the sense of an RG flow. Terms with consecutive creation or annihilation operators on the same site are prohibited because each site is singly-occupied in the ground state. It is now easy to cast this Hamiltonian into a more recognizable form. For \(\sigma^\prime=\sigma\), we get
\begin{equation}\begin{aligned}
	\sum_{\sigma}\delta_{\sigma,\sigma^\prime}c^\dagger_{1\sigma}c_{2\sigma}c^\dagger_{2\sigma^\prime}c_{1\sigma^\prime} = \sum_\sigma \left(n_{1\sigma} - n_{1\sigma}n_{2\sigma}\right) ~,
\end{aligned}\end{equation}
while \(\sigma=-\sigma^\prime=\pm 1\) gives 
\begin{equation}\begin{aligned}
	\sum_{\sigma}\delta_{\sigma,-\sigma^\prime}c^\dagger_{1\sigma}c_{2\sigma}c^\dagger_{2\sigma^\prime}c_{1\sigma^\prime} = -\left(S_1^+ S_2^- + \text{h.c.}\right)~.
\end{aligned}\end{equation}
For the latter expression, we introduced the local spin-flip operators \(S_i^\pm\). The expression above it can also be cast into spin variables, using the equations
\begin{equation}\begin{aligned}
	\frac{1}{2}\sum_{\sigma}n_{i\sigma} = \frac{1}{2},\\
    \frac{1}{2}\sum_{\sigma}\sigma n_{i\sigma} = S_i^z,
\end{aligned}\end{equation}
where the first equation is simply the condition of half-filling at each site, and the second equation is the definition of the local spin operator in \(z-\)direction. Adding and subtracting the equations gives \(n_{i\sigma} = \frac{1}{2} + \sigma S_i^z\)~.

Substituting everything back into eq.~\ref{effHam1} and dropping constant terms gives
\begin{equation}\begin{aligned}
	H_\text{eff} = 2\frac{V_H^2}{U/2}\left(2S_1^z S_2^z + S_1^+S_2^- + S_1^-S_2^+\right) = J_\text{eff} {\bf S}_1\cdot{\bf S}_2~,
\end{aligned}\end{equation}
where the effective antiferromagnetic Heisenberg coupling is \(J_\text{eff} = \frac{8V_H^2}{U}\).

\end{document}